\documentclass[review]{elsarticle}

\usepackage{color} 
\usepackage{amsmath} 

\usepackage{lineno,hyperref}
\modulolinenumbers[5]

\begin{document}

\begin{frontmatter}

\title{PIBM: Particulate immersed boundary method for fluid-particle interaction problems}


\author[mymainaddress]{Hao Zhang}

\author[mymainaddress]{F. Xavier Trias}
\author[mymainaddress]{Assensi Oliva\corref{mycorrespondingauthor1}}
\cortext[mycorrespondingauthor1]{Corresponding author1}
\ead{cttc@cttc.upc.edu}
\author[myforthaddress]{Dongmin Yang\corref{mycorrespondingauthor2}}
\cortext[mycorrespondingauthor2]{Corresponding author2}
\ead{d.yang@leeds.ac.uk}
\author[mysecondaryaddress]{Yuanqiang Tan}
\author[mythirdaddress]{Shi Shu}
\author[myforthaddress]{Yong Sheng}

\address[mymainaddress]{Heat and Mass Transfer Technological Center, Technical University of Catalonia,
 Terrassa, Barcelona 08222, Spain}
 \address[myforthaddress]{School of Civil Engineering, University of Leeds, Leeds LS2 9JT, UK}
\address[mysecondaryaddress]{ School of Mechanical Engineering, Xiangtan University, Hunan 411105, China}
\address[mythirdaddress]{ School of Mathematics and Computational Science, Xiangtan University, Hunan 411105, China}

\begin{abstract}

It is well known that the number of particles should be scaled up to enable industrial scale simulation.
The calculations are more computationally intensive when the motion of the surrounding fluid is considered. Besides the 
advances in computer hardware and numerical algorithms, the coupling scheme also plays an important role on the computational 
efficiency. In this study, a particulate immersed boundary method (PIBM) for simulating the fluid-particle multiphase flow was 
presented and assessed in both two- and three-dimensional applications. The idea behind PIBM derives from the 
conventional momentum exchange-based immersed boundary method (IBM) by treating each Lagrangian point as a solid particle.
This treatment enables LBM to be coupled with fine particles residing within a particular grid cell. 
Compared with the conventional IBM, dozens of times speedup in two-dimensional simulation and hundreds of times in 
three-dimensional simulation can be expected under the same particle and mesh number. Numerical simulations of particle 
sedimentation in Newtonian flows were conducted based on a combined 
lattice Boltzmann method - particulate immersed boundary method - discrete element method scheme, showing that the PIBM 
can capture the feature of particulate flows in fluid and is indeed a promising 
scheme for the solution of the fluid-particle interaction problems.  

\end{abstract}


\end{frontmatter}


\section{Introduction}

Due to the stochastic nature of the solid particle behaviors, the fluid-particle interaction problems are 
often too complex to be solved analytically or observed by physical experiments. Therefore, 
they have to be analyzed by means of numerical simulations. In our previous work~\cite{ZHANGCAF2014}, 
we have reported a numerical study of particle sedimentation process by using a combined 
Lattice Boltzmann Method~\cite{qian1992lattice}, Immersed Boundary 
Method~\cite{Peskin1977220} and Discrete Element Method~\cite{hao8} (LBM-IBM-DEM) scheme. 
The LBM-IBM-DEM scheme is attractive because no artificial parameters are required in the calculation of 
both fluid-particle and particle-particle interaction force. However, the computational cost of this coupling 
scheme not only lies 
on the grid resolutions in LBM and the solid particle number $N_{P}$, but also highly depends on the number of the 
Lagrangian points $N_{LP}$ distributed on the solid particle boundaries. Since $N_{LP}$ on each particle 
should be large enough to ensure the accurate calculation of the fluid-particle interaction force and torque, the actual 
point number considered in the numerical interpolation is $N_{P}\times N_{LP}$ which makes the main calculation effort 
in the LBM-IBM-DEM modeling highly related to the IBM part. For the system of two-dimensional 504 particles with
each particle containing 57 Lagrangian points~\cite{ZHANGCAF2014}, a calculating period of one month may be needed to 
simulate the entire sedimentation process in a $2~cm\times 2~cm$ cavity in a single CPU without additional parallel 
accelerations such as the graphics processing unit 
(GPU)~\cite{yue2013} or Message Passing Interface (MPI)~\cite{kafui2}. This computational efficiency is significantly 
lower than other coupling schemes based on the Navier-Stokes equations and DEM (NS-DEM)~\cite{hao46,Tan2012137,Zhang2012467} 
when treating the same amount of solid particles. The bottleneck of LBM-IBM-DEM 
scheme becomes dramatically serious in the three-dimensional applications. The important feature of the coupled NS-DEM 
simulations is that one single fluid cell can contain several solid particles, and the fluid-particle interaction force is 
calculated based on the local porosity in the cell together with the superficial slip velocity between particle and 
fluid~\cite{difelice}. In the NS-DEM simulations, the details of particle geometry are not considered when the size of 
the particles are significantly smaller than the system characteristic scale. Alternatively, the LBM-DEM simulations tell
a different story in which each solid particle is constructed by dozens of lattice units (or more in three-dimensional cases) 
and the hydrodynamics force acting on each particle is 
the resultant of forces on the Lagrangian points and obtained by integrating around the circumference of the solid 
particle~\cite{Cook2004,Han20071080,zhangcsse2008,cui2014coupled}. Although the latter coupling scheme seems to be more 
rational, it is highly limited by the current computational capability as also argued by Zhu et al.~\cite{zhu2007discrete} in
their review paper and thus simulations of industrial scale problems are not computationally affordable. Yu and Xu~\cite{yu2003particle} stated that: 
``At this stage of development the difficulty in particle-fluid flow modeling is mainly related to the solid phase rather 
than the fluid phase.'' A numerical method that can be widely accepted in engineering application is the one with superior 
computational convenience. This paper aims at improving the computational efficiency of our previous LBM-IBM-DEM 
scheme~\cite{ZHANGCAF2014} and extending the coupling scheme to three-dimensional cases. The idea of the traditional 
NS-DEM is borrowed here to treat each Lagrangian point directly as one solid 
particle, therefore, one single LBM grid is allowed to contain several solid particles spatially.  

The available works on LBM-DEM were reviewed in~\cite{ZHANGCAF2014} where the calculation of fluid-particle interaction 
force is regarded as the key point and it requires an accurate description of the boundaries of 
the solid particles. In general, there are two ways to do this, namely the Immersed Moving Boundary method (IMB) proposed by 
Noble and Torczynski~\cite{doi:10.1142/S0129183198001084} and the IBM proposed by Peskin~\cite{Peskin1977220}.
Here, we focus on the IB-LBM simulation. Feng and Michaelides firstly proposed a penalty IB-LBM scheme \cite{Feng2004602} 
and then improved it via a direct forcing scheme~\cite{Feng200520}. Instead, Niu et al.~\cite{Niu2006173} proposed a simpler,
parameter-free and more efficient momentum exchange-based IB-LBM. The scheme of Niu et al.~\cite{Niu2006173}
has been inherited by numerous researchers to study the Fluid-Structure Interaction (FSI) problems~\cite{FLD:FLD2023,yuan2014momentum}, 
thermal flows~\cite{Wang201498Thermal,hu2013natural} and particulate flows~\cite{wu2010,ZHANGCAF2014} due to its natural advantage.
In this study, the fluid-particle interaction force is also evaluated by the scheme of Niu et al.~\cite{Niu2006173} without 
introducing any artificial parameters. Unlike the aforementioned treatments in which the Lagrangian points were linked by 
stable solid bonds~\cite{wu2010,ZHANGCAF2014} or flexible filaments~\cite{yuan2014momentum}, the constraints between the 
Lagrangian points are thoroughly removed. By doing so, the free floating of the Lagrangian points is allowed and the 
driving force on them is simply based on the momentum exchange of the fluid particles. Hereby, the new coupling scheme 
is called Particulate Immersed Boundary Method (PIBM) to show the difference to Niu et al.~\cite{Niu2006173}.
It is worthwhile mentioning that Wang et al.~\cite{wang2013lattice} carried out a coupled LBM-DEM simulation to study the 
gas-solid fluidization in which the size of the particles is smaller than the lattice spacing, and the Energy-Minimization 
Multi-Scale (EMMS)~\cite{li1994particle} drag model is adopted to calculate the coupling force between solid and gas phase. 
However, Wang et al.~\cite{wang2013lattice} only conducted two-dimensional simulations and the establishment of an empirical 
formula containing the local porosity is still needed. In addition, the EMMS has a lower computational performance than
the direct momentum exchange-based scheme as adopted in current study.

The rest of the paper is organized as follow. To make this paper self-contained, the mathematics of the 
three-dimensional LBM, PIBM and DEM were briefly introduced in Section~\ref{LBMPIBMDEM}. 
In Section~\ref{Numericalresults}, case studies of the particle sedimentation 
in Newtonian flow were presented with the numerical results discussed. Finally, some conclusions were made in 
Section \ref{conclusion}.  

\section{Governing equations}\label{LBMPIBMDEM}

\subsection{Lattice Boltzmann model with single-relaxation time collision}

 \begin{figure}[!h]
 \centering
 \includegraphics[width=0.5\textwidth]{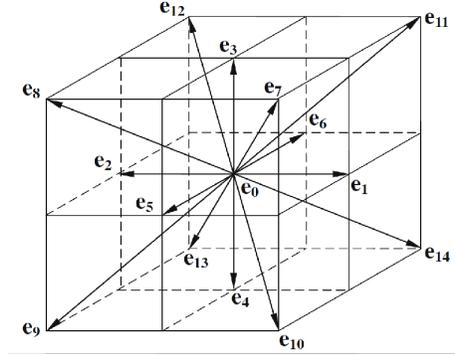}
 \vskip-0.2cm
 \caption{Schematic diagram of the D3Q15 model\cite{Wu20105022}.} \label{PIBMsch}
 \end{figure}

We consider the simulation of the incompressible Newtonian fluids where the LBM-D3Q15 model~\cite{qian1992lattice} is adopted, 
and the spatial distribution of the velocities is shown in Figure~\ref{PIBMsch}.
Following the same notation used by Wu and Shu\cite{Wu20105022}, those 15 lattice velocities are given by

\begin{eqnarray}
e_{\alpha} = 
\left\{
\begin{array}{ll}
   (0,0,0)c & \alpha=0 \\
   (\pm 1,0,0)c, (0,\pm 1,0)c, (0,0,\pm 1)c & \alpha=1-6  \\
   (\pm 1,\pm 1,\pm 1)c & \alpha = 7-14 
\end{array}
\right. 
\end{eqnarray}

\noindent where $c$ is termed by the lattice speed. The formulation of the lattice Bhatnagar-Gross-Krook model is

\begin{equation}\label{lbm2}
f_{\alpha}(r+e_{\alpha}\delta_{t},t+\delta_{t})=f_{\alpha}(r,t)-\frac{f_{\alpha}(r,t)-f_{\alpha}^{eq}(r,t)}{\tau}+F_{b}\delta_{t}
\end{equation}

\noindent where $f_{\alpha}(r,t)$ represents the fluid density distribution function, $r=(x,y,z)$ stands for the space
position vector, $t$ denotes time and $\tau$ denotes the non-dimensional relaxation time,
$F_{b}\delta_{t}$ denotes the fluid-solid interaction force term which is given in the following section.
 The equilibrium density distribution function, $f_{\alpha}^{eq}(r,t)$, can be written as

\begin{equation}\label{lbm3}
f_{\alpha}^{eq}(r,t)= \rho_{f} \omega_{\alpha} [1+3(e_{\alpha}\cdot u)+\frac{9}{2}(e_{\alpha}\cdot u)^{2}-\frac{3}{2}\mid u \mid^{2}]
\end{equation}

\noindent where the value of weights are: $\omega_{0}=2/9$, $\omega_{\alpha}=1/9$ for $\alpha=1-6$ and 
$\omega_{\alpha}=1/72$ for $\alpha=7-14$. $u$
denotes the macro velocity at each lattice node which can be calculated by 
$u=(\sum\limits_{\alpha=0}^{14}f_{\alpha}e_{\alpha})/\rho_{f}$, and the macro fluid density is obtained by 
$\rho_{f}=\sum\limits_{\alpha=0}^{14}f_{\alpha}$.

\subsection{ Particulate immersed boundary method (PIBM)}

 \begin{figure}[!h]
 \centering
 \includegraphics[width=0.54\textwidth]{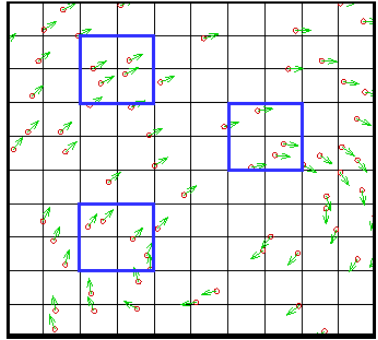}
 \vskip-0.2cm
 \caption{Schematic diagram of the PIBM.} \label{PIBM}
 \end{figure}

For the sake of clarity, the two-dimensional schematic diagram of the PIBM is given in Figure~\ref{PIBM} followed by 
three-dimensional equation systems.
As shown, the fluid is described using the Eulerian square lattices and the solid particles are denoted by the Lagrangian 
points moving in the flow field. Instead of using several Lagrangian points to 
construct one large solid particle~\cite{ZHANGCAF2014}, each Lagrangian point is treated as one single solid particle 
in this study. The fluid density distribution functions on the solid particles are evaluated using the numerical 
extrapolation from the circumambient fluid points, 

\begin{equation}\label{lbm5}
f_{\alpha}(X_{l},t)=L \cdot f_{\alpha}(r,t)
\end{equation}

\noindent where $X_{l}(X,Y,Z)$ is the coordinates of the solid particles, $L$ is the three-dimensional 
Lagrangian interpolated polynomials, 

\begin{equation}\label{lbm6}
L=\sum\limits_{ijk}\left(\prod\limits_{l=1,l!=i}^{i_{max}}\frac{X-x_{ljk}}{x_{ijk}-x_{ljk}}\right)
\left(\prod\limits_{m=1,m!=j}^{j_{max}}\frac{Y-y_{imk}}{y_{ijk}-y_{imk}}\right)
\left(\prod\limits_{n=1,n!=k}^{k_{max}}\frac{Z-z_{ijn}}{z_{ijk}-z_{ijn}}\right)
\end{equation}

\noindent where $i_{max}$, $j_{max}$  and $k_{max}$ are the maximum numbers of the Eulerian points used 
in the extrapolation as shown by blocks in Figure~\ref{PIBM}. With the movement of the solid particle, 
$f_{\alpha}(X_{l},t)$ 
will be further affected by the particle velocity, $U_{p}$,

\begin{equation}\label{lbm7}
f_{\beta}(X_{l},t+\delta_{t})=f_{\alpha}(X_{l},t)-2\omega_{\alpha}\rho_{f}\frac{e_{\alpha}U_{p}}{c_{s}^{2}}
\end{equation}

\noindent where the subscript $\beta$ represents the opposite direction of $\alpha$. Based on the momentum exchange 
between fluid and particles, the force density, $g(X_{l},t)$, at each solid particle can be calculated 
using $f_{\alpha}$ and $f_{\beta}$,

\begin{equation}\label{lbm8}
g(X_{l},t)=\sum\limits_{\beta}e_{\beta}[f_{\beta}(X_{l},t)-f_{\alpha}(X_{l},t)]
\end{equation}

The effect on the flow fields from the solid boundary is the body force term $F_{b}\delta_{t}$ in Equation~\ref{lbm2}, 
where $F_{b}$ can be expressed by

\begin{equation}\label{lbm9}
F_{b}=\left(1-\frac{1}{2\tau}\right)\omega_{\alpha}\left(3\frac{e_{\alpha}-u}{c^{2}}+9\frac{e_{\alpha}\cdot u}{c^{4}}e_{\alpha}\right)F(r,t)
\end{equation}

\noindent and

\begin{equation}\label{lbm10}
F(r,t)=\sum\limits_{l}g(X_{l},t)D_{ijk}(r_{ijk}-X_{l})A_{p}
\end{equation}

\noindent $A_{p}$ is the cross-sectional area of the particle which is given as $A_{p}=0.25\pi d_{p}^{2}$, $d_{p}$ 
is the diameter of the particle. $D_{ijk}$ is used to restrict the feedback force to only take
 effect on the neighbor of interface and is given by

\begin{equation}\label{lbm11}
D_{ijk}(r_{ijk}-X_{l})=\frac{1}{h^{3}}\delta_{h}\left(\frac{x_{ijk}-X_{l}}{h}\right)\delta_{h}\left(\frac{y_{ijk}-Y_{l}}{h}\right)\delta_{h}\left(\frac{z_{ijk}-Z_{l}}{h}\right)
\end{equation}

\noindent with

\begin{equation}\label{lbm12}
  \delta_{h}(a)= \left\{
   \begin{array}{cc}
   \frac{1}{4}(1+cos(\frac{\pi a}{2})), &  when \mid a \mid \leq 2  \\
   0, & otherwise \\
   \end{array}
   \right.
  \end{equation}

\noindent where $h$ is the mesh spacing. It should be stressed that by adding a body force on the flow field, the
macro moment flux also has to be modified by the force $\rho_{f} u=\sum\limits_{\alpha=0}^{14}f_{\alpha}e_{\alpha}+\frac{1}{2}F(r,t)\delta t$.

On the other hand, the fluid-solid interaction force exerted on the solid particle can
be obtained as the reaction force of $g(X_{l},t)$,

\begin{equation}\label{lbm13}
F_{fpi}=-g(X_{l},t) A_{p}
\end{equation}

\subsection{Modeling of the particle-particle interactions}

 The dynamic equations of the particle can be expressed as
\begin{eqnarray}\label{demgov}
m\frac{d^{2}r}{d t^{2}} &=& (1-\frac{\rho_{f}}{\rho_{p}})g+F_{fpi} \\
I\frac{d^{2}\theta}{d t^{2}} &=& \tau
\end{eqnarray}
where $m$ and $I$ are the mass and the moment of inertia of the particle, respectively.
 $r$ is the particle position and $\theta$ is the angular position. $\rho_{f}$ and $\rho_{p}$
 are the densities of the fluid and particle, respectively. $g$ is the gravitational acceleration and
  $\tau$ is the torque. Considered forces on the right hand side of Equation~\ref{demgov} are 
 the buoyant force and the fluid-particle interaction
 force $F_{fpi}$. When the particles collide directly with other particles or the walls, the DEM~\cite{hao8} is employed to 
calculate the collision force. In this study, the particles and walls are directly specified by material 
properties in the simulation such as density, Young's modulus and friction coefficient.  
When the collisions take place, the theory of Hertz~\cite{Johnson} is used for modeling the force-displacement 
relationship while the theory of Mindlin and Deresiewicz~\cite{MD} is employed for the tangential force-displacement 
calculations. For two particles of radius $R_{i}$ , Young's modulus $E_{i}$
and Poisson's ratios $\nu_{i}$
$(i=1,2)$, the normal force-displacement relationship reads

\begin{eqnarray}
F_{n}=\frac{4}{3}E^{*}R^{*1/2}\delta_{n}^{3/2}
\end{eqnarray}

\noindent where the equivalent Young's modulus and radius can be calculated by 
$1/E^{*}=(1-\nu_{1}^{2})/E_{1}+(1-\nu_{2}^{2})/E_{2}$ and
$1/R^{*}=1/R_{1}+1/R_{2}$, respectively.

The incremental tangential force arising from an incremental tangential displacement depends on the
loading history as well as the normal force and is given by

\begin{eqnarray}
\Delta T=8G^{*}r_{a}\theta_{k}\Delta \delta_{t}+(-1)^{k}\mu\Delta F_{n}(1-\theta_{k})
\end{eqnarray}

\noindent where $1/G^{*}=(1-\nu_{1}^{2})/G_{1}+(1-\nu_{2}^{2})/G_{2}$, $r_{a}=\sqrt{\delta_{n}R^{*}}$ is radius of the contact area. $\Delta \delta_{t}$ is the relative tangential 
incremental surface displacement, $\mu$ is the coefficient of friction, the value of $k$ and $\theta_{k}$ changes with the loading
history. 

\section{Results and discussions}\label{Numericalresults}

\begin{table}
\centering
\begin{tabular}{rrrr}
  &  \\ 
\hline
Solid phase &  & Fluid phase &     \\ 
\hline
Density ($kg\cdot m^{-3}$)      & 1010 & Density ($kg\cdot m^{-3}$)      & 1000\\   
Young's Module ($G Pa$)      & 68.95 & Viscosity ($kg\cdot m^{-1}\cdot s^{-1}$)      & 1.0e-3\\   
Poisson ratio ($N\cdot m^{-1}$)      & 0.33 & Lattice length ($m$)      & 0.0001\\ 
Friction coefficient ($-$)      & 0.33 & Gravity acceleration ($m\cdot s^{-2}$)      & 9.8\\ 
\hline
\end{tabular}
 \caption{Properties of the particles and fluid.}
\label{Properties}
 \end{table}
 
As stated in previous section, comparing with the conventional IBM, several essential simplifications have been made 
in the PIBM including removing the constraints between the Lagrangian particles and omitting the calculation of hydrodynamics 
torque. A natural question is that can the PIBM still success in the complex fluid-particle interaction problems with frequent
momentum transfer? For the sake of demonstrating the capability of the PIBM, two- and three-dimensional simulations of 
particle sedimentation in Newtonian liquid in a cavity were carried out. This configuration is interesting because the 
Rayleigh-Taylor instability phenomenon may take place on the interface of the agglomerating particles and the fluid. 
In two-dimensional case~\cite{ZHANGCAF2014}, the fluid in the lower half of the cavity is found to insert into the upper 
half and this forms a fluid pocket of mushroom shape in the particle phase interior. Then, the relative smooth interface 
between the two phase is disturbed and the fluid pocket is teared to small ones. These fluid pockets have the appearance 
of irregular shape and travel at both vertical and horizontal speed until all the particles fall down on the cavity bottom. 
In this study, the two-dimensional results by PIBM were directly given due to the fact that the collision rule of the D2Q9 
model is very similar to D3Q15~\cite{qian1992lattice} and the two-dimensional code has been tested in ~\cite{ZHANGCAF2014}. 
In the rest of this section, the accuracy of the PIBM was firstly examined by simulating the falling process of a single 
particle in Newtonian flow and the results were compared with the analytical solutions based on the Stokes' law. By means of 
the comparison, the parameters were also calibrated and adopted in the following multi-particle simulations. Then, the two- 
and three-dimensional results were presented in Section~\ref{Sedimentation2D} and ~\ref{Sedimentation3D}, respectively. 
The physical properties of the particles and the surrounding fluid are given in Table~\ref{Properties}. It should be mentioned
that the lattice spacing length, $h$, is $0.0001 m$ in all the simulations. 

\subsection{Falling of a single particle}\label{Effectgrid}

\begin{figure}[!ht]
\centering
\includegraphics[angle= -90,width=0.475\textwidth]{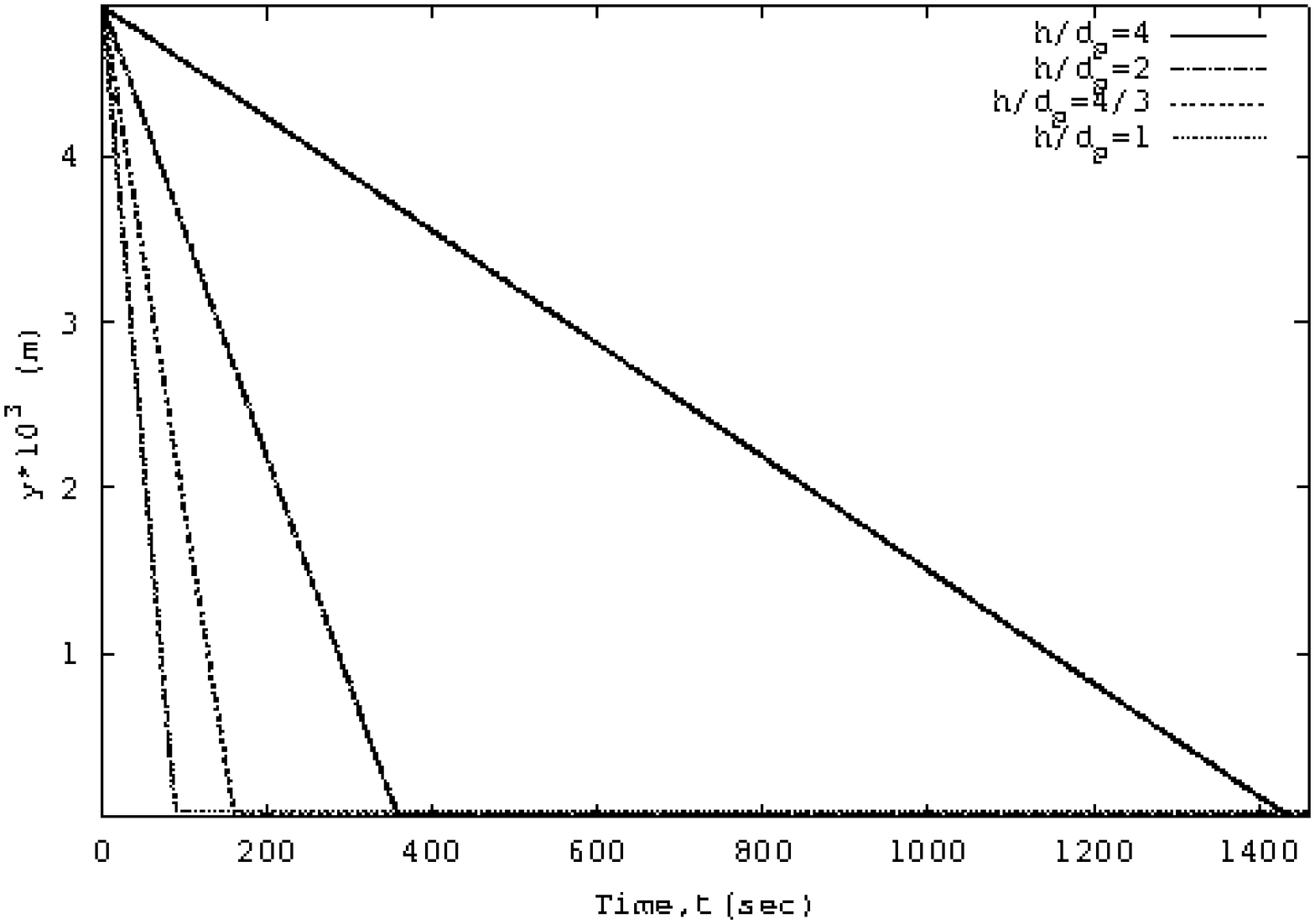}
\hspace{2mm}
\includegraphics[angle= -90,width=0.475\textwidth]{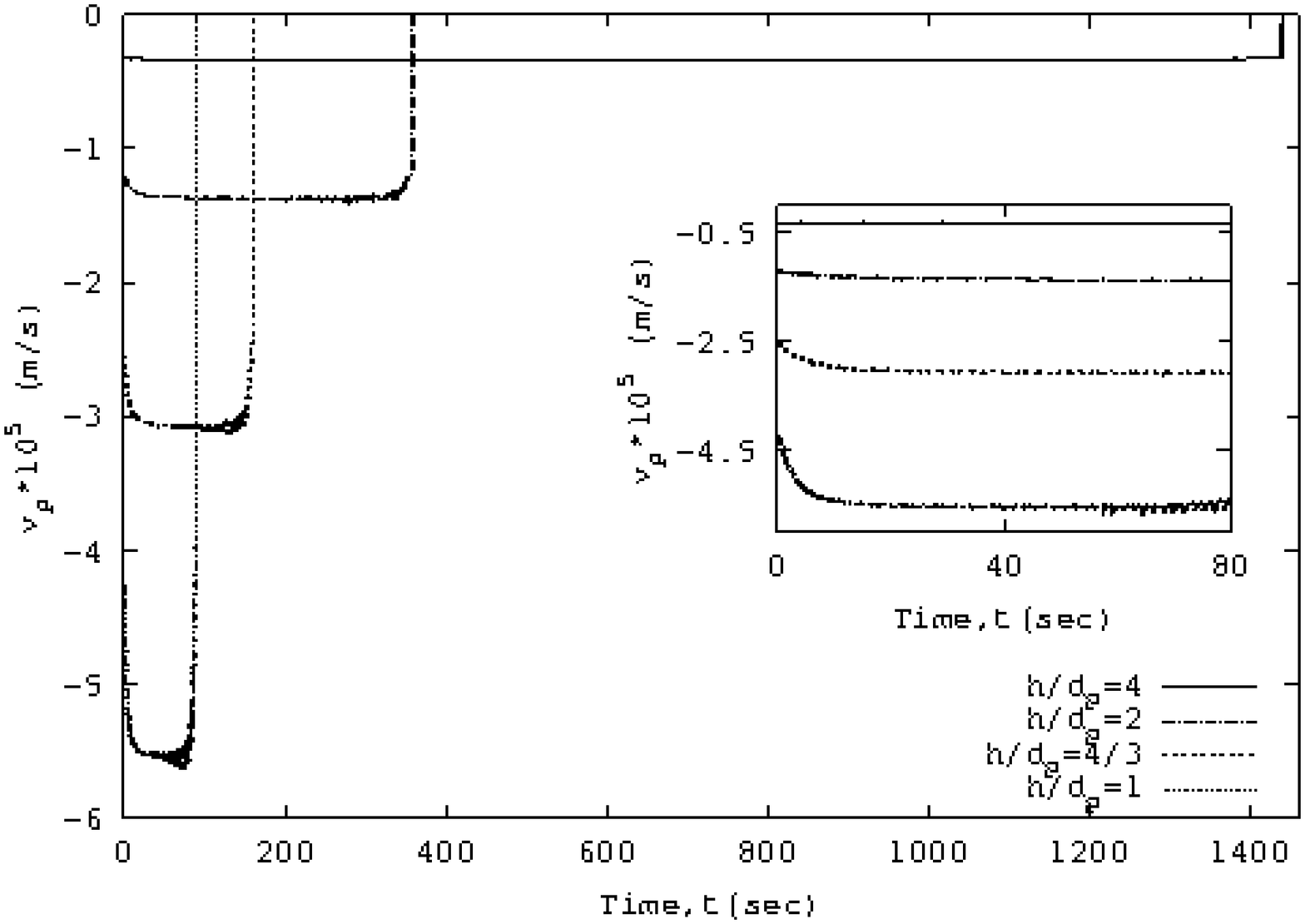}
\\
(a) \hspace{52mm} (b)
\caption{(a) longitudinal coordinates and (b) longitudinal velocities of the particle under different $h/d_{p}$.}
\label{settling}
\end{figure}

\begin{figure}
\centering
\includegraphics[angle=  0,width=0.3\textwidth]{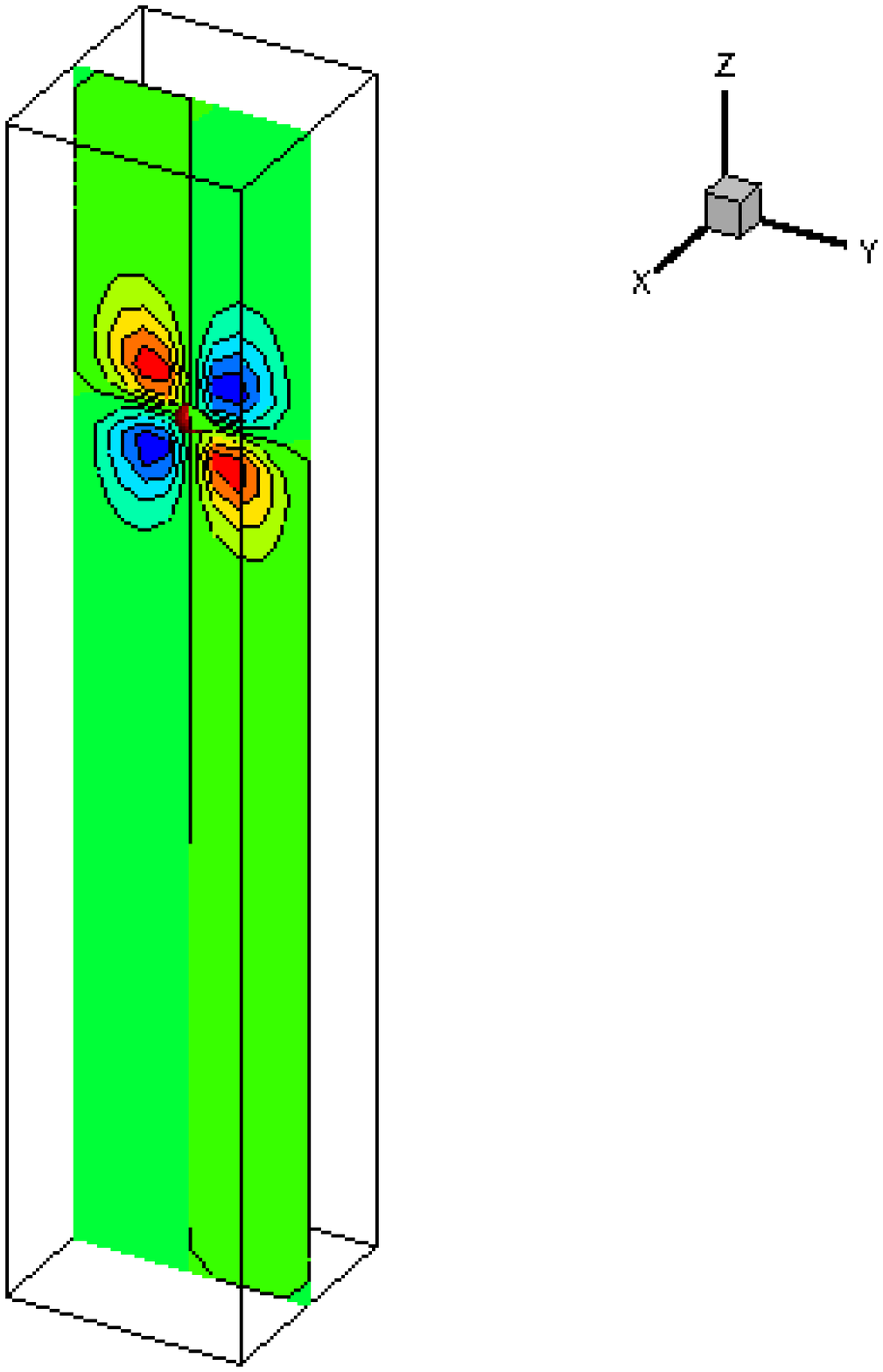}
\hspace{20mm}
\includegraphics[angle=  0,width=0.3\textwidth]{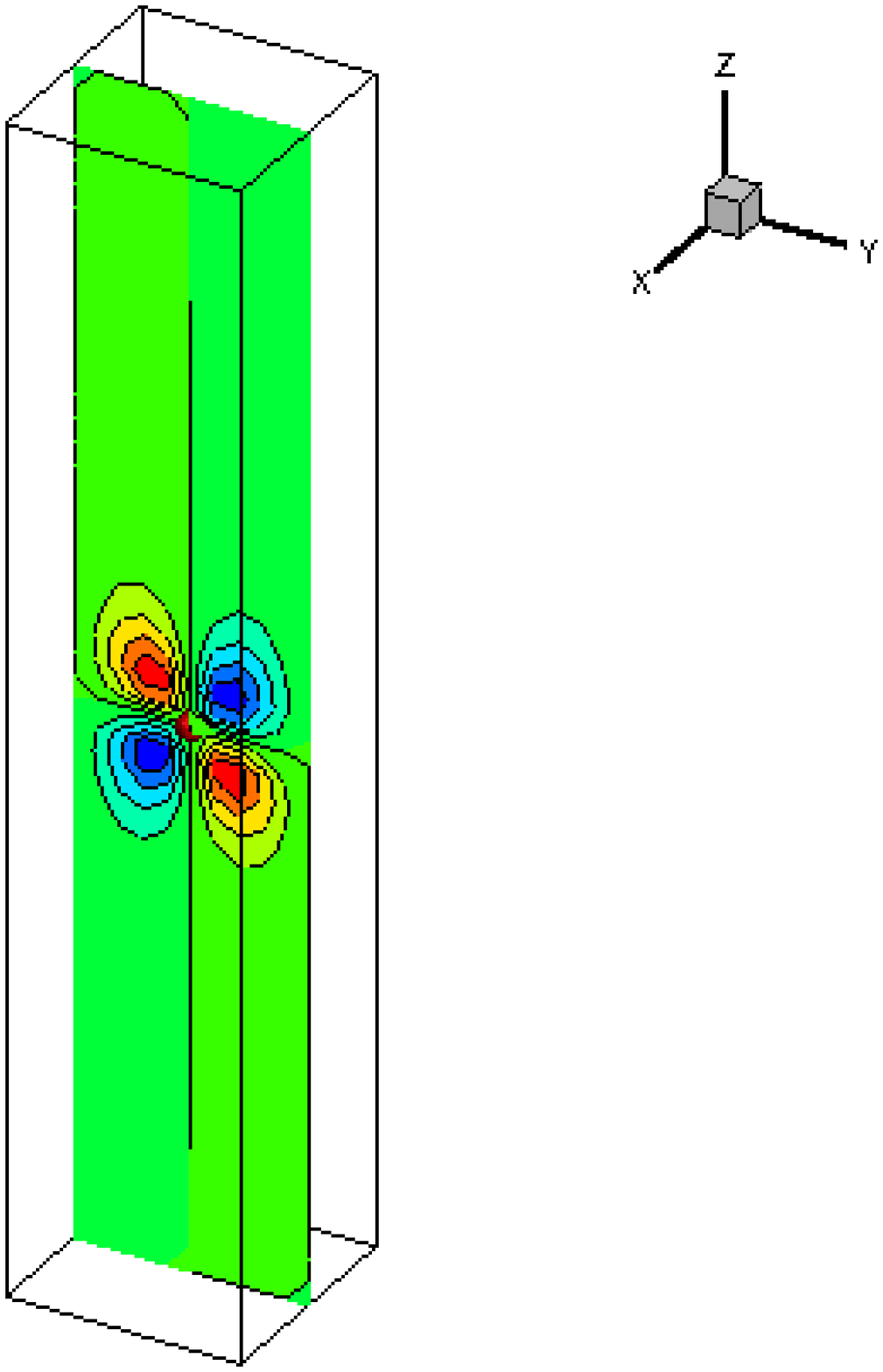}
\\
(a) \hspace{52mm} (b)
\\
\includegraphics[angle=  0,width=0.3\textwidth]{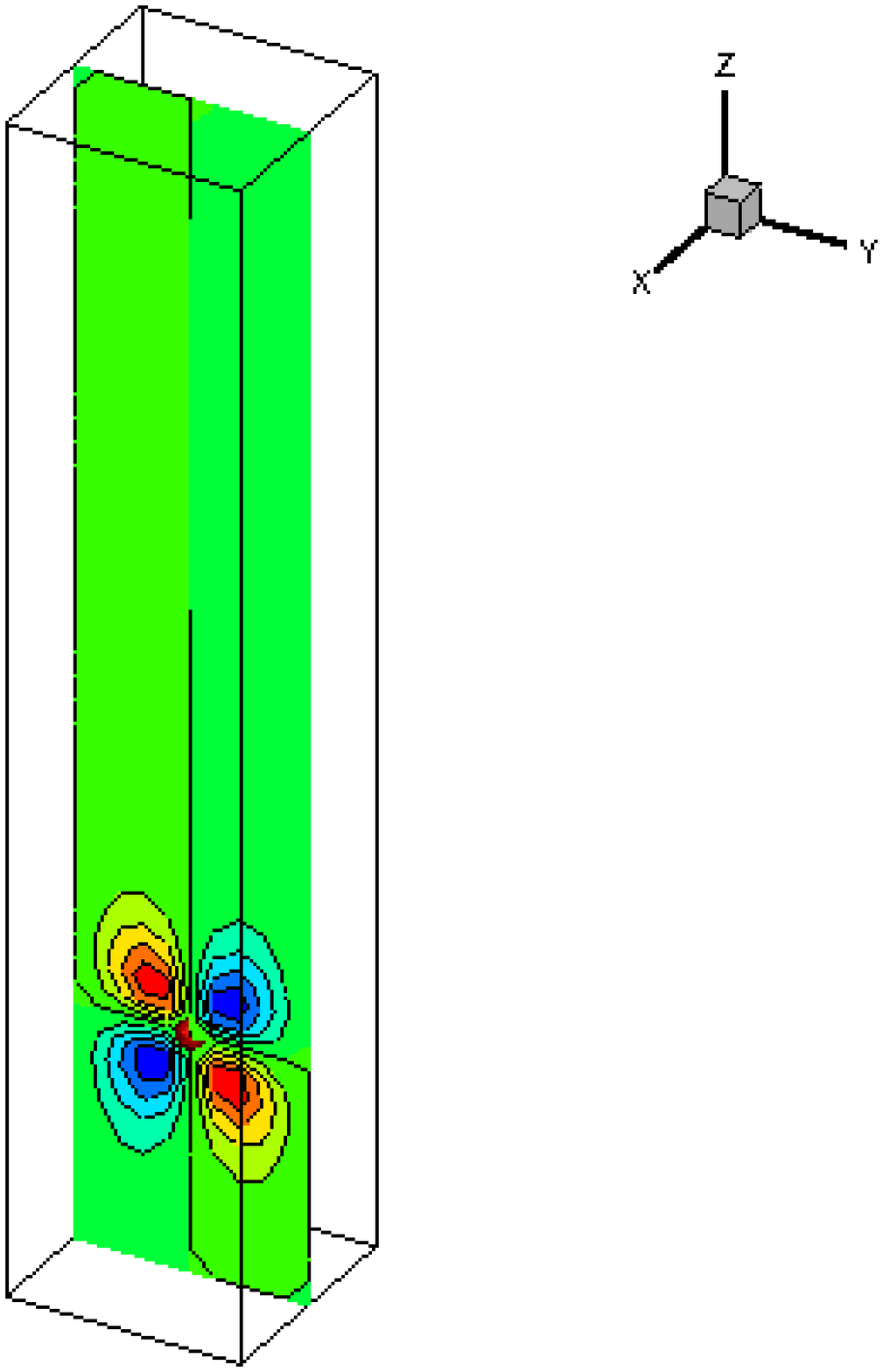}
\hspace{20mm}
\includegraphics[angle=  0,width=0.3\textwidth]{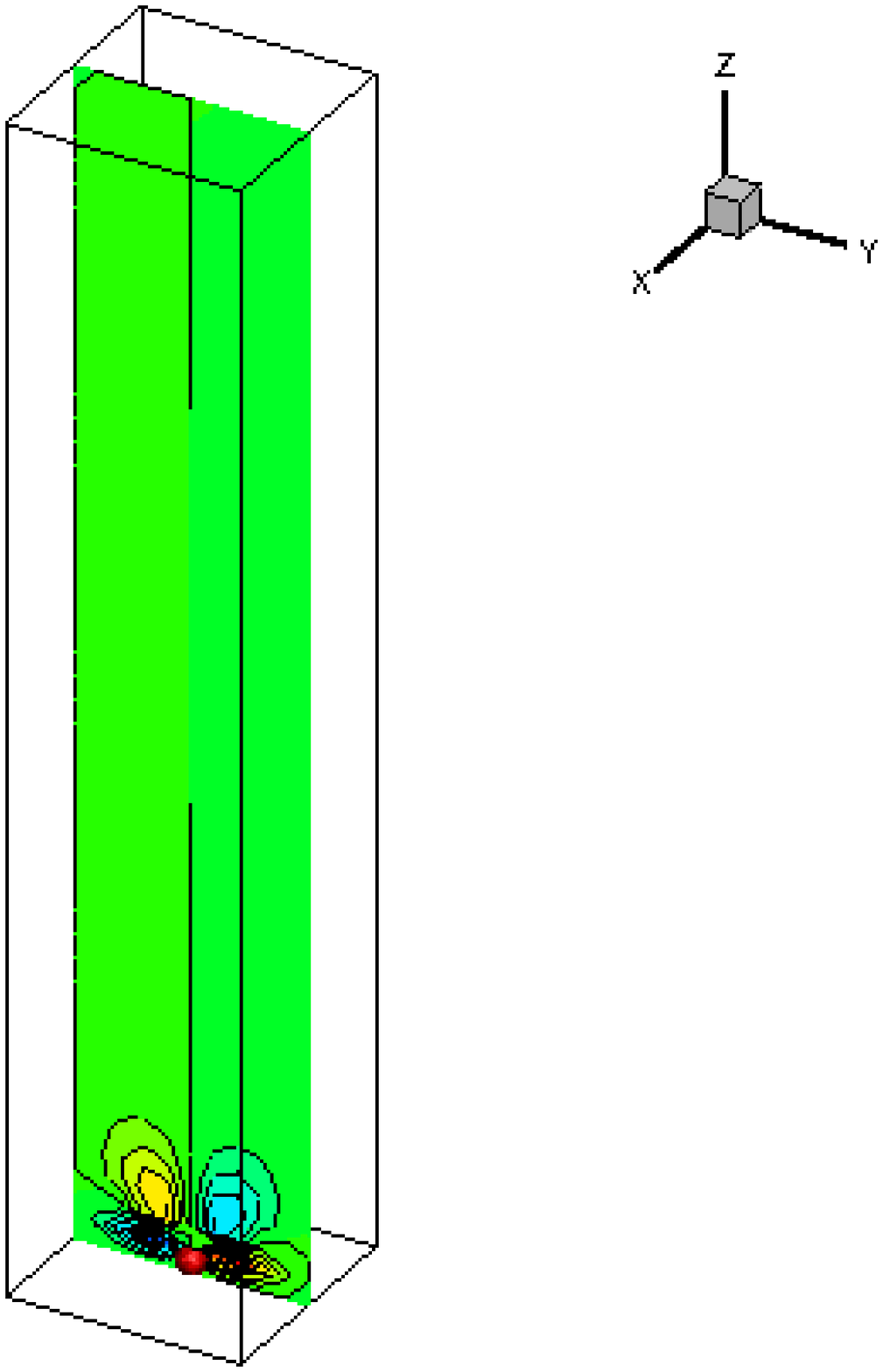}
\\
(c) \hspace{52mm} (d)
\caption{Instantaneous height of the $0.0001 m$ particle with contour plots for $v$ at time 
(a) $t=22.5 s$, (b) $t=45.0 s$, (c) $t=67.5 s$, (d) $t=90.0 s$.}
\label{Instantaneousheight}
\end{figure}

Falling of the single particle in a cuboid cavity was firstly investigated.
The length and width of the cuboid cavity are $0.001 m$ and the height is $0.005 m$. The initial position of the 
particle is at $(0.0005m,0.0005m,0.0049m)$. Four kinds of particles with different diameters are considered, 
namely $d_{p}=0.25\times10^{-4} m$, $0.5\times10^{-4} m$, $0.75\times10^{-4} m$ 
and $10^{-4} m$ or $h/d_{p}=4$, $2$, $4/3$ and $1$, respectively. 
The largest diameter is equal to one LBM grid spacing length. The longitudinal coordinates and 
velocities of different particles during the falling process are shown in Figure~\ref{settling}. The particles 
at rest begin to deposit under the effect of the gravitational force. After a period of acceleration, 
the particles fall with a constant settling velocity until they 
approaches to the bottom. The magnitude of the settling velocity increases with the particle size. Finally, the 
particles stay at the bottom of the cavity with zero longitudinal velocity. Figure~\ref{Instantaneousheight} shows 
several typical snapshots of the falling process of the $1\times10^{-4} m$ particle with contour plots for $v_{f}$, 
clear influence of the particle motion on the flow structure can be observed. For dilute suspensions, the settling 
velocity of a single particle in a viscous fluid flow can be evaluated by the Stokes' law which is given by

\begin{equation}\label{lbm14}
 V_{s} = \dfrac{(\rho_{p}-\rho_{f})d_{p}^{2}g}{18\mu}
\end{equation}

\noindent where $\mu$ is the dynamic viscosity of the fluid. 
Quantitative comparison between the results based on the Stokes' law and the numerical ones are presented in 
Table~\ref{Stokes}. As shown, the settling velocities predicted by numerical simulation agree well with the 
Stokes' law. However, it is found that the particle may oscillate around the center line during the falling process
and the fluctuation on the velocity increases with the particle size especially closing to the cavity bottom. 
Feng et al.~\cite{feng2014using} also reported this unsteadiness phenomenon using coupled Direct Numerical 
Simulation and DEM (DNS-DEM). In the following subsections of this study, $h/d_{p}=4$ and $2$ were chosen based on 
the similar criterion 
as adopted in the NS-DEM simulations~\cite{kafui2002discrete}. Our numerical simulations show that this 
ratio works well in the multi-particle cases in general, however, further numerical and experimental validations may 
be needed to fully assess its effect on the particle behaviors.

\begin{table}[!ht]
\centering
\begin{tabular}{ccccc}
  &  \\ 
\hline
$h/d_{p}$ &  Based on Stokes' law  & Numerical results & $\tau$ & Physical timestep    \\
          &    ($m/s$)  &   ($m/s$) &   &  ($s$)    \\
          
\hline
$4$    & $-3.40\times10^{-6}$ & $-3.41\times10^{-6}$  & $0.65$ & 0.0005 \\   
$2$    & $-1.36\times10^{-5}$ & $-1.37\times10^{-5}$   & $0.72$ & 0.0007\\ 
$4/3$    & $-3.06\times10^{-5}$ & $-3.07\times10^{-5}$  & $0.79$ & 0.0010 \\ 
$1$    & $-5.44\times10^{-5}$ & $-5.52\times10^{-5}$   & $0.85$ & 0.0012 \\ 
\hline
\end{tabular}
 \caption{The settling velocities at different particle size.}
\label{Stokes}
 \end{table}


\subsection{Sedimentation of two-dimensional particles in Newtonian flows}\label{Sedimentation2D}

\begin{figure}[!ht]
\centering
\includegraphics[angle=  0,width=0.45\textwidth]{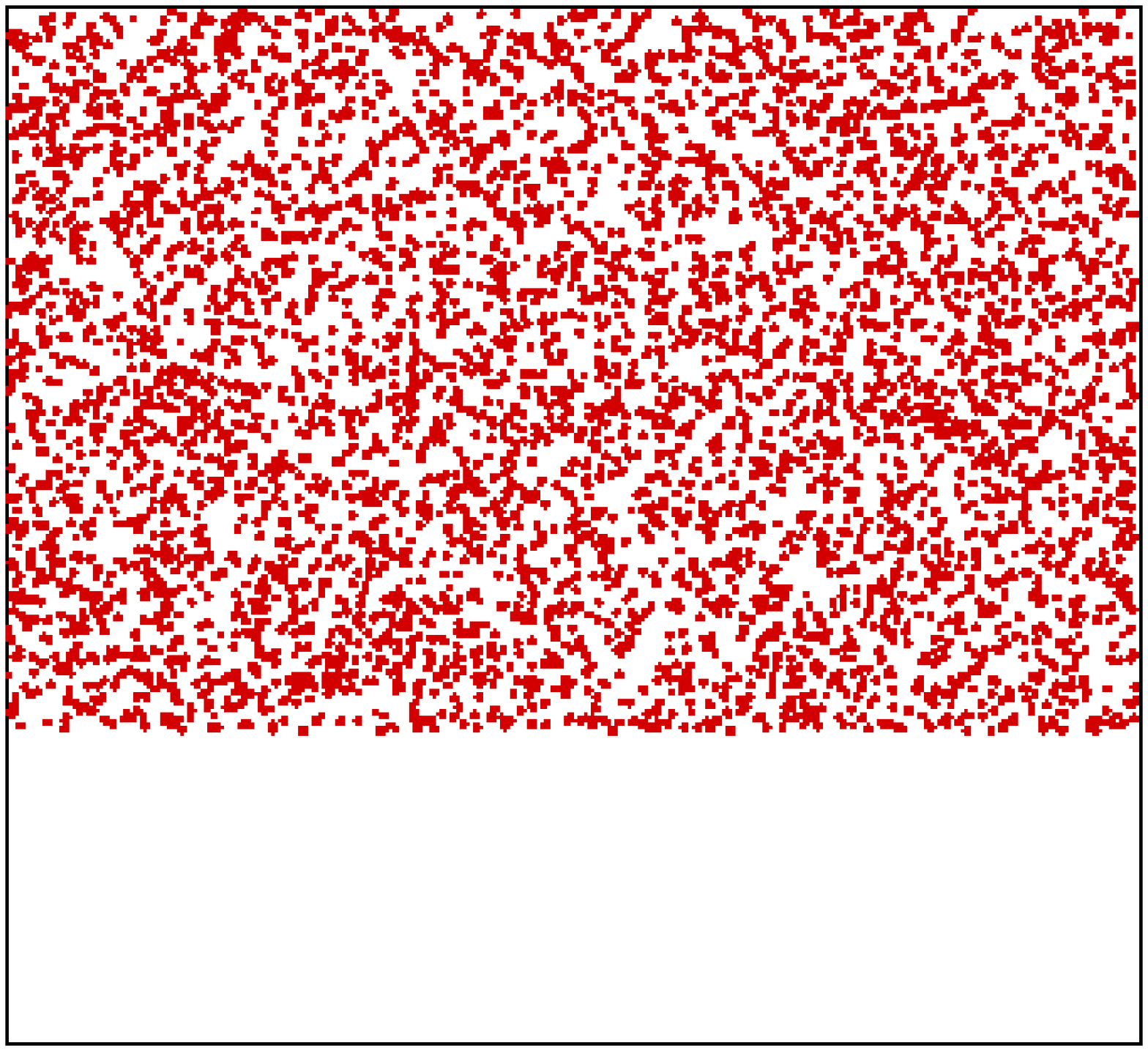}
\hspace{2mm}
\includegraphics[angle=  0,width=0.45\textwidth]{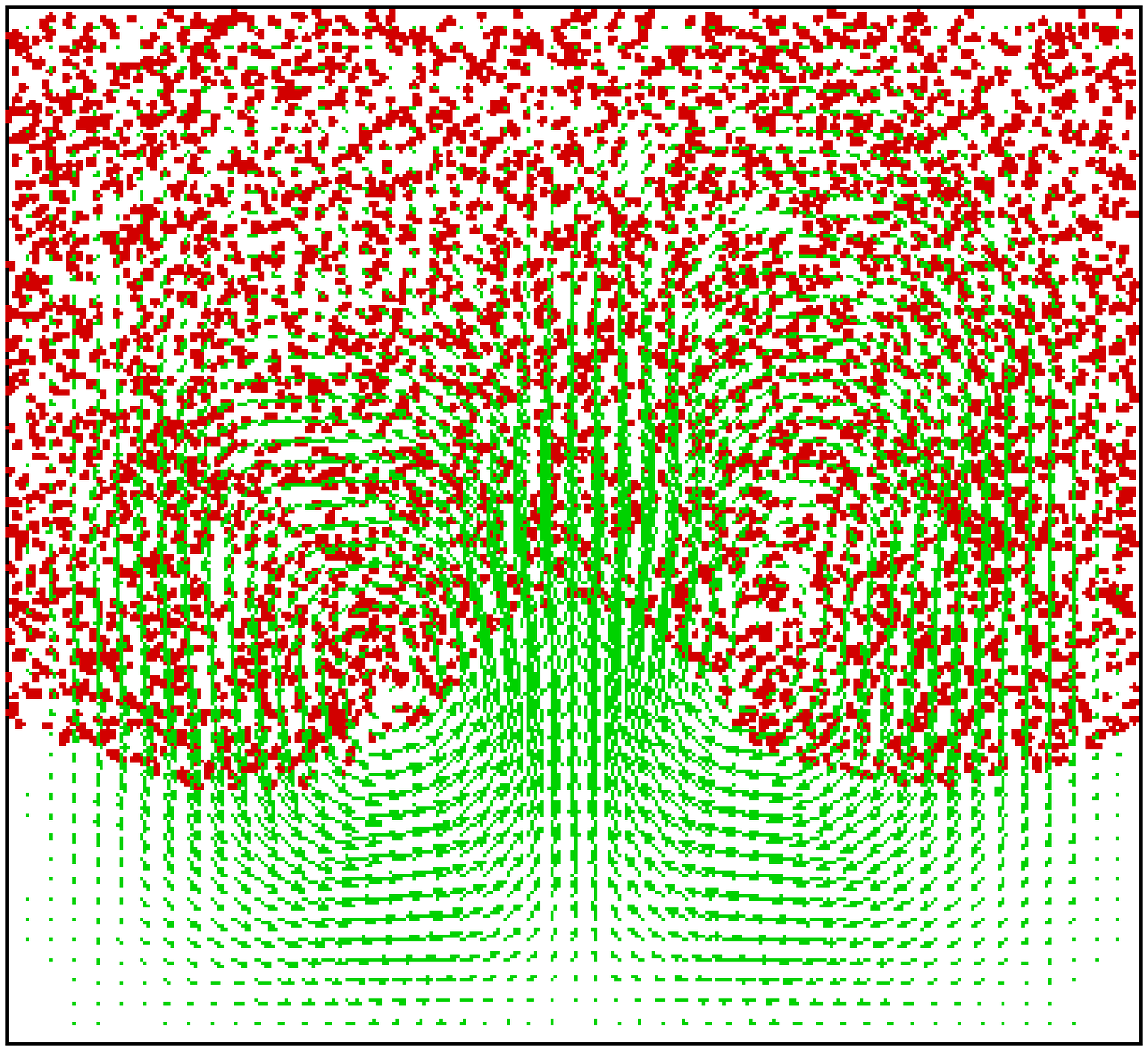}
\\
(a) \hspace{52mm} (b)
\\
\includegraphics[angle=  0,width=0.45\textwidth]{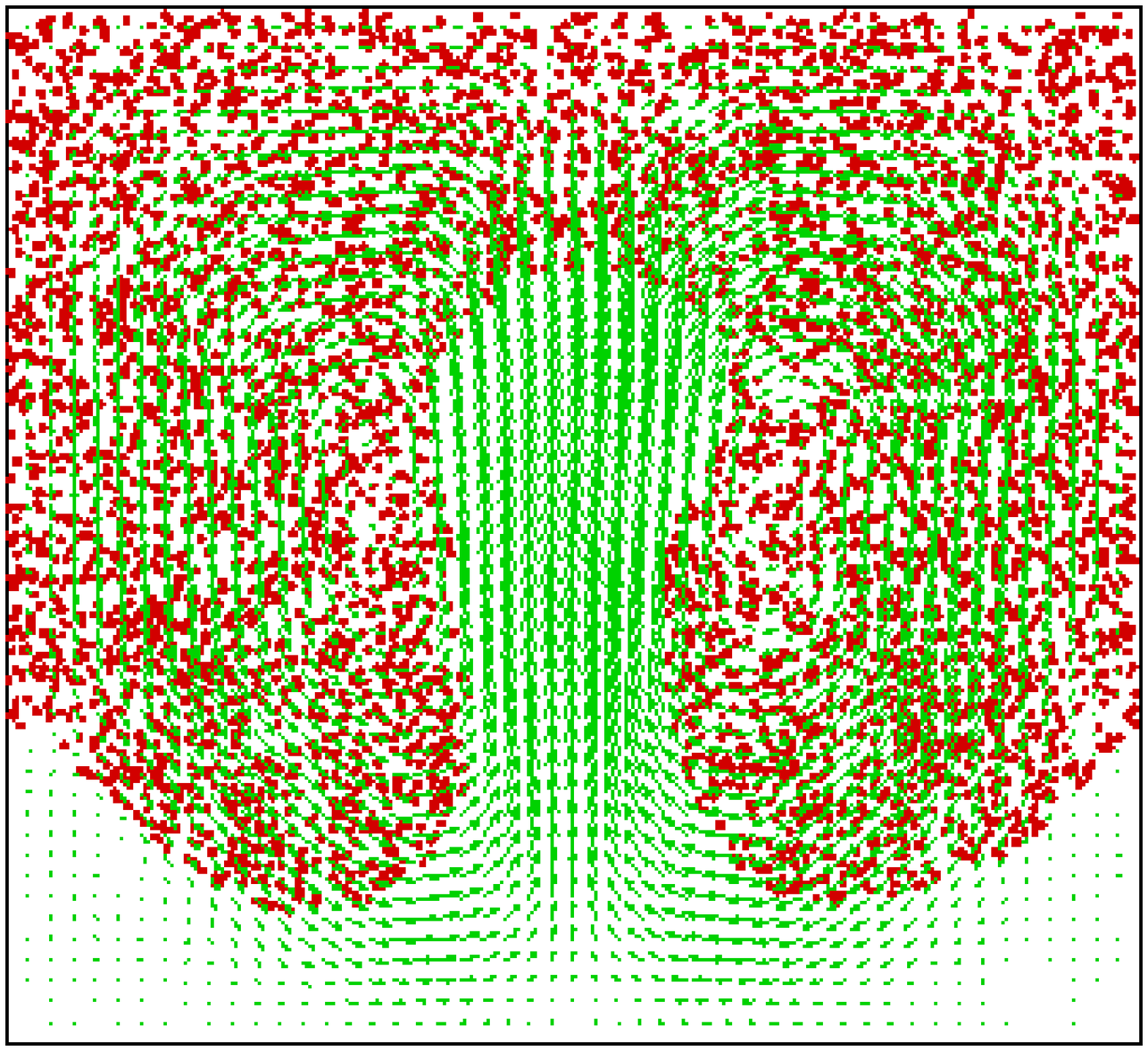}
\hspace{2mm}
\includegraphics[angle=  0,width=0.45\textwidth]{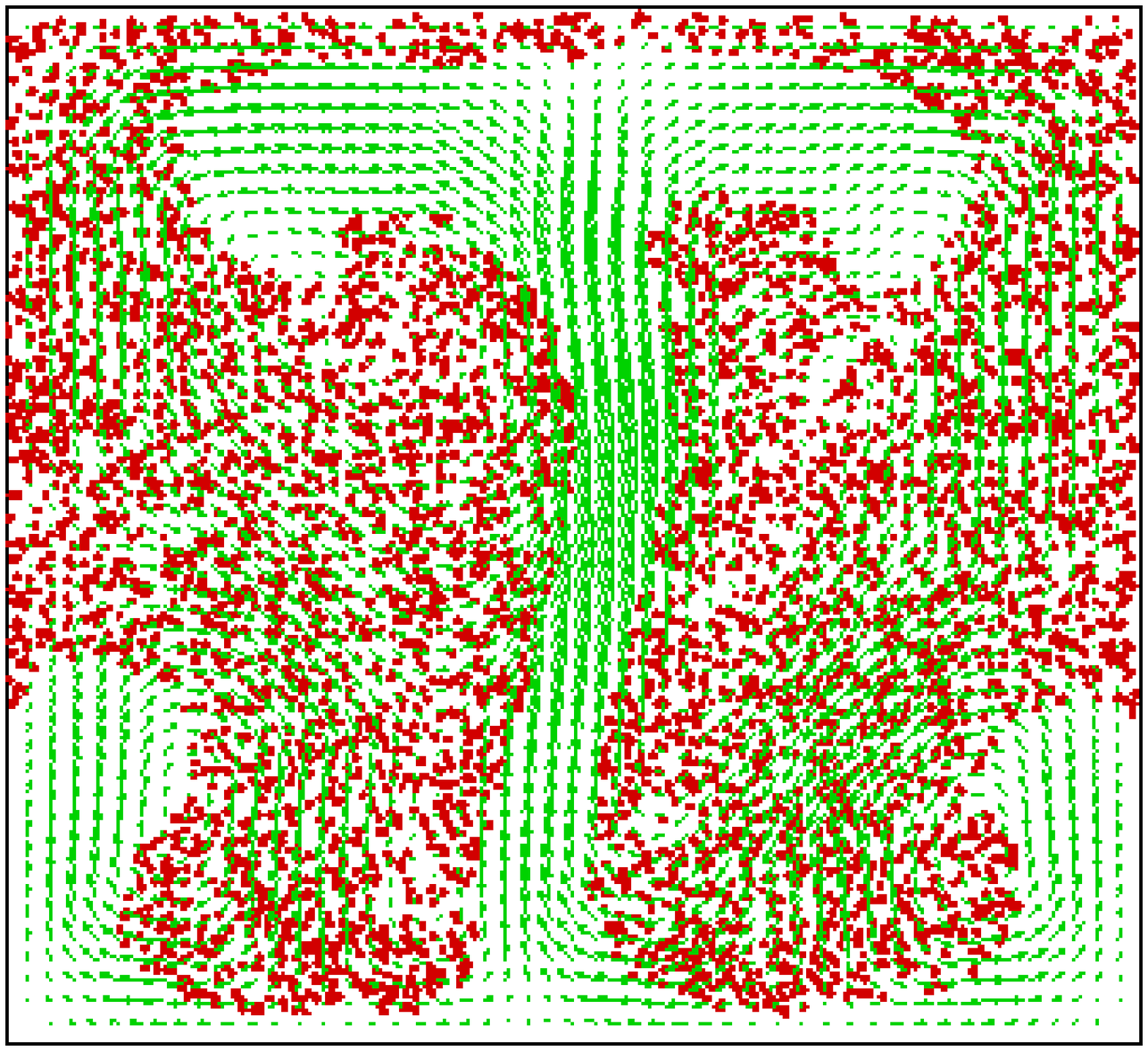}
\\
(c) \hspace{52mm} (d)
\caption{Instantaneous particle distribution with the fluid velocities at time 
(a) $t=0.0 s$, (b) $t=2.5 s$, (c) $t=5.0 s$, (d) $t=10.0 s$.}
\label{5k2d1zao}
\end{figure}

\begin{figure}[!ht]
\centering
\includegraphics[angle=  0,width=0.45\textwidth]{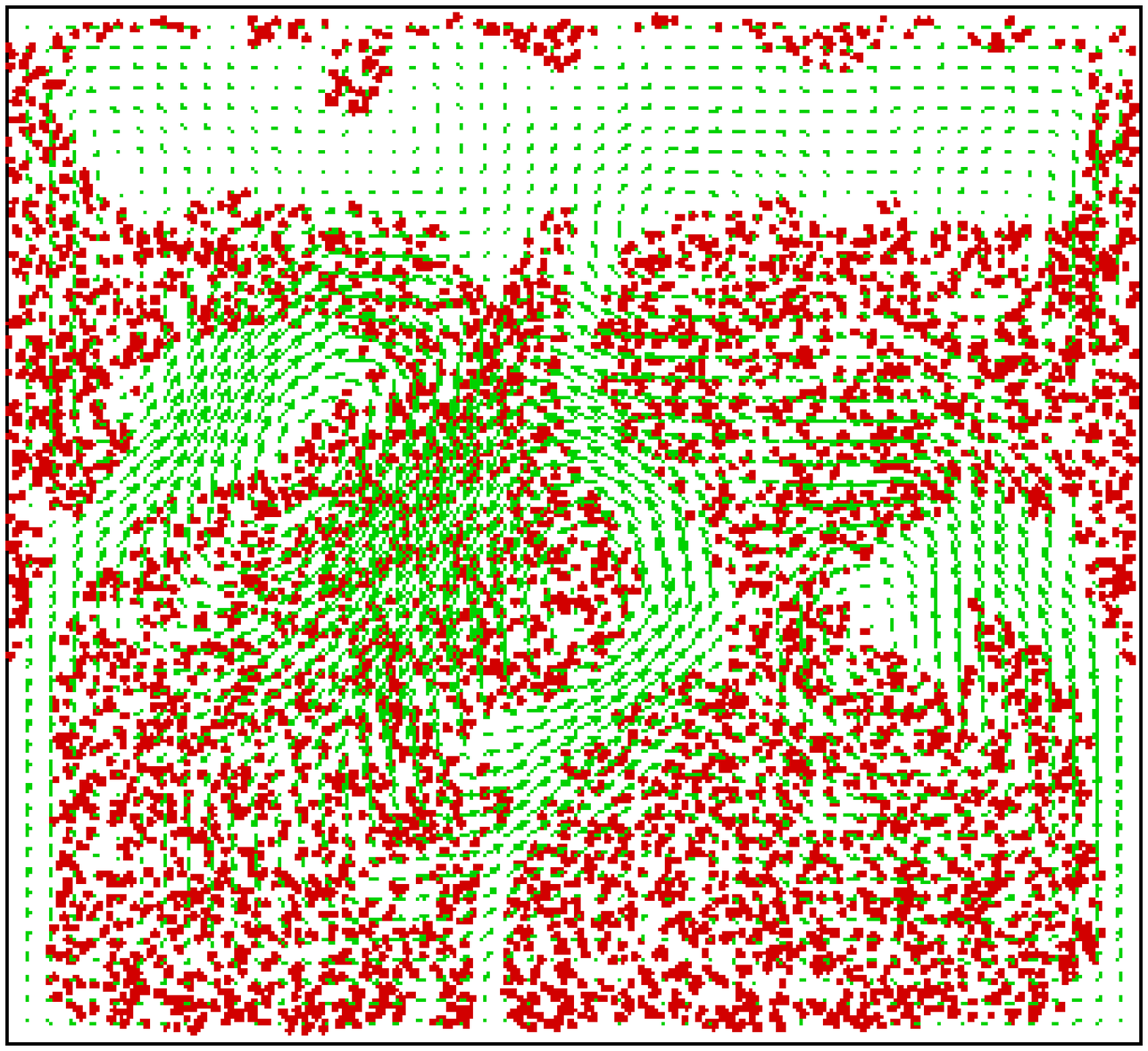}
\hspace{2mm}
\includegraphics[angle=  0,width=0.45\textwidth]{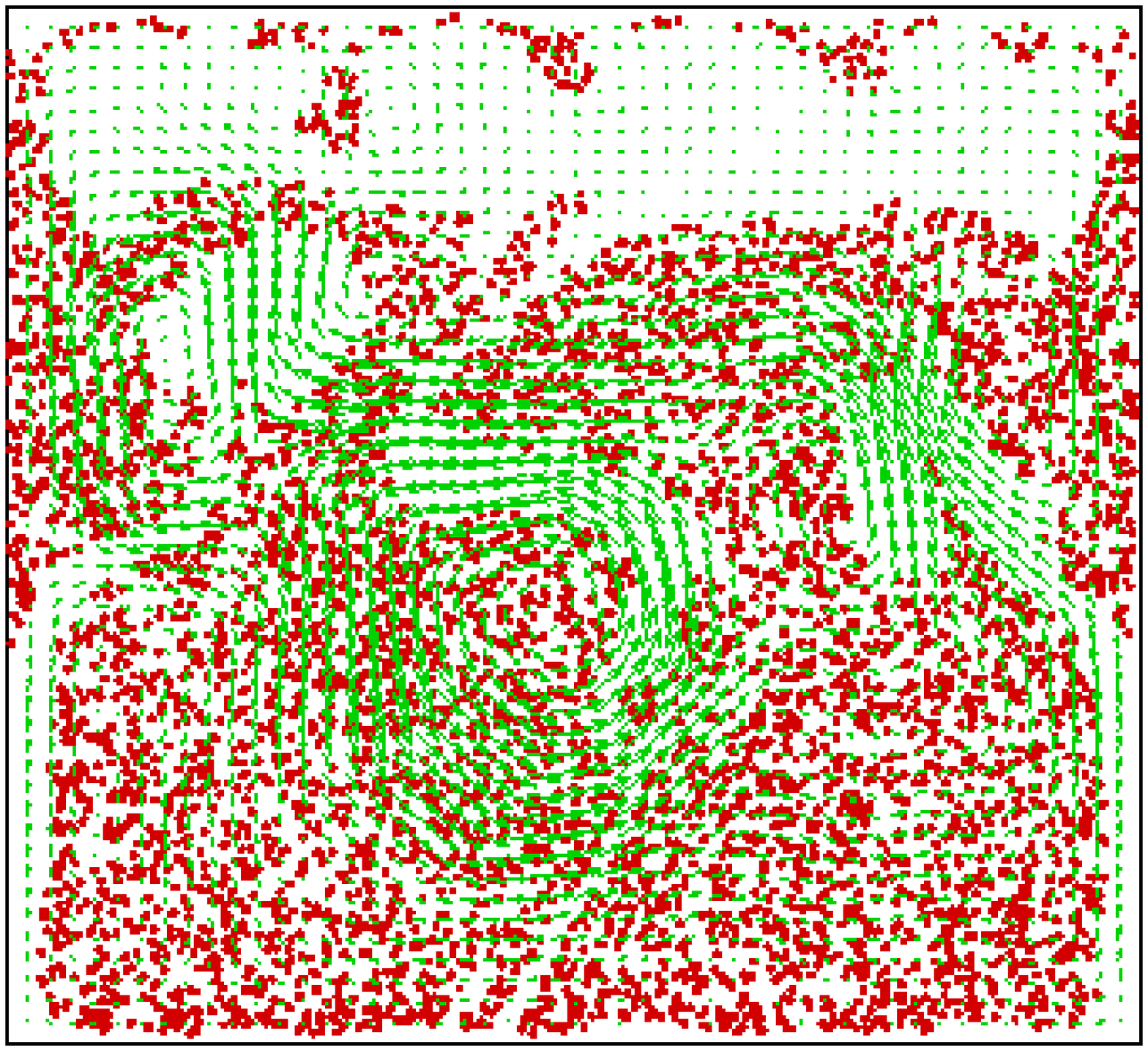}
\\
(a) \hspace{52mm} (b)
\\
\includegraphics[angle=  0,width=0.45\textwidth]{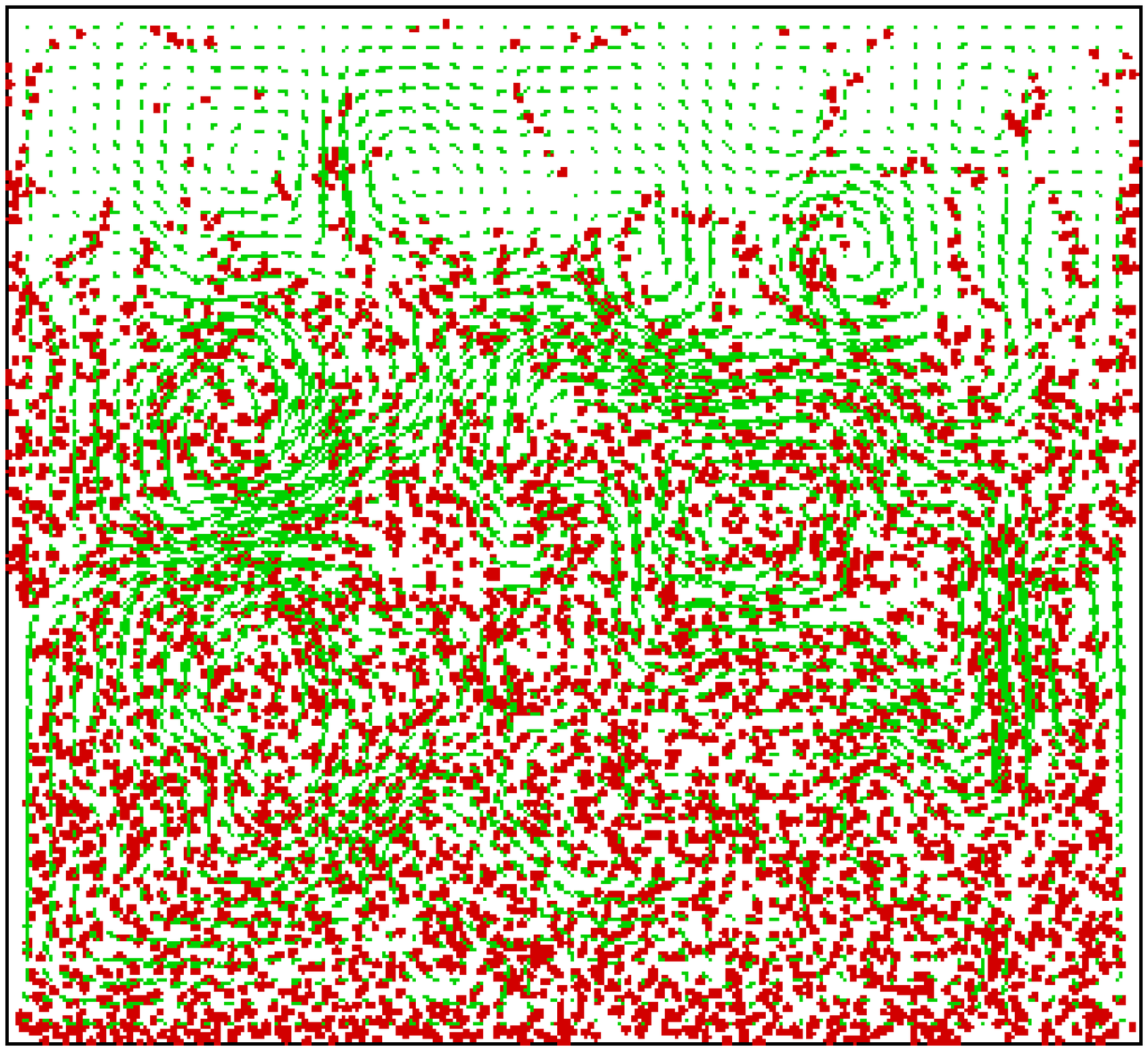}
\hspace{2mm}
\includegraphics[angle=  0,width=0.45\textwidth]{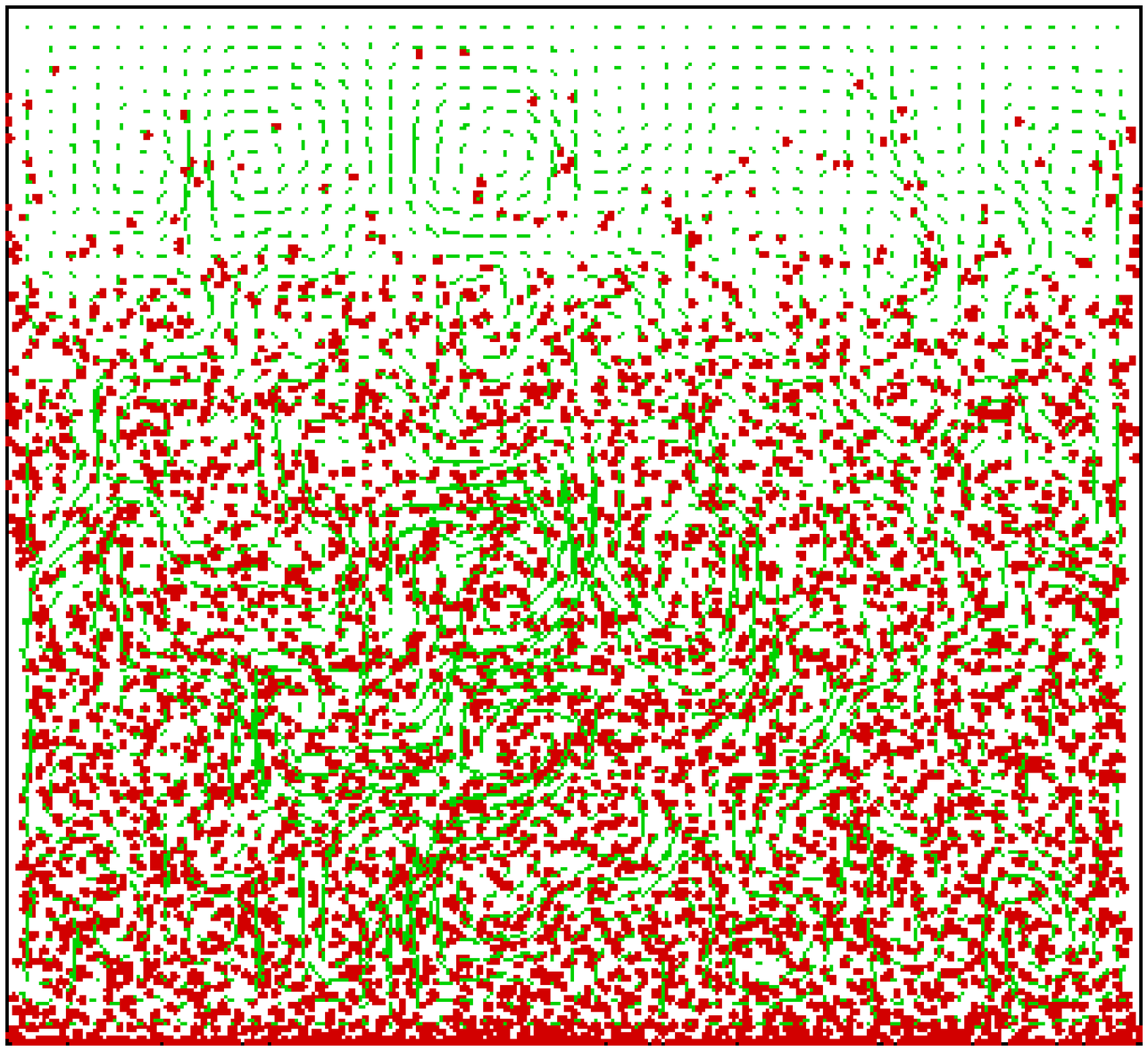}
\\
(c) \hspace{52mm} (d)
\caption{Instantaneous particle distribution with the fluid velocities at time 
(a) $t=20.0 s$, (b) $t=25.0 s$, (c) $t=50.0 s$, (d) $t=100.0 s$.}
\label{5k2d1wan}
\end{figure}

Two-dimensional simulations of the particle sedimentation in a square cavity have been conducted using various 
numerical methods~\cite{glowinski1999distributed,Feng2004602,Feng200520,ZHANGCAF2014}. Here we consider a 
$0.01 m \times 0.01 m$ cavity with 5000 two-dimensional particles. The properties of the particles and the 
surrounding fluid are given in Table~\ref{Properties}. The diameter of the particles are $0.25\times10^{-4} m$ 
or $h/d_{p}=4$. The relaxation time is  $\tau=0.65$, it leads to a physical timestep of $0.0005 s$. Initially, 
the 5000 particles are randomly generated in the upper three-fifths domain and then deposit under the effect of the 
 gravitational force. Figure~\ref{5k2d1zao} displays the changing process of the interface line from straight to curve.
 As expected, the fluid at the lower half of the cavity is swallowed into the the agglomerating particles forming a 
 open hole of mushroom shape. The open hole is shattered to pieces when the particles fall down as shown in Figure~\ref{5k2d1wan}.
 Generally speaking, the patterns observed in this simulation are very close to the results provided in the former 
 references~\cite{glowinski1999distributed,Feng2004602,Feng200520,ZHANGCAF2014}. However, compared with the results 
 of large particles that calculated using conventional momentum exchange-based immersed boundary 
 method~\cite{ZHANGCAF2014}, the whole sedimentation process takes much longer time due to the low settling velocity.
 
\subsection{Sedimentation of three-dimensional particles in Newtonian flows}\label{Sedimentation3D}

\subsubsection{The sedimentation process}

\begin{figure}[!ht]
\centering
\includegraphics[angle=  0,width=0.56\textwidth]{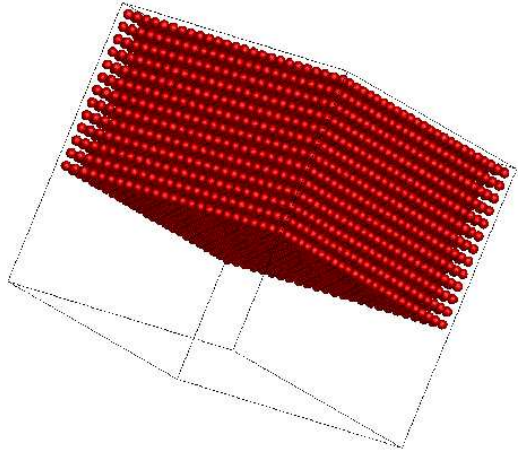}
\caption{Positions of the 8125 particles at time $t=0.0 s$.}
\label{8125initial}
\end{figure}

In this subsection, a three-dimensional $0.0015 m \times 0.00015 m \times 0.0015 m$ cubic cavity is considered. 
The diameter of the particles is $0.5\times10^{-4} m$ or $h/d_{p}=2$. 
The relaxation time is  $\tau=0.72$ which leads to a physical timestep of $0.0007 s$. Initially,
$8125$ particles are positioned in the upper three-fifths domain as shown in Figure~\ref{8125initial}, 
the solid fraction is $0.15$, total volume occupied by the particle assembly is $1.9\times10^{-9} m^{3}$, 
total volume of the particles is $5.3\times10^{-10} m^{3}$ and thus the local porosity is $0.719$.
There are vertically 13 layers of particles, in each layer there are $625$ particles forming a $25\times25$ matrix.
In each direction, the particles are uniformly distributed. The gap between the horizontal neighboring 
particles and between the closest particles and the side wall is about $0.00001 m$. The gap between the 
vertical neighboring particles and between the highest particles and the top wall is about $0.000018 m$. 
The no-slip boundary is adopted on the six boundaries of the cavity, namely the fluid nearby the wall 
will have zero velocity.

\begin{figure}[!ht]
\centering
\includegraphics[angle=  0,width=0.47\textwidth]{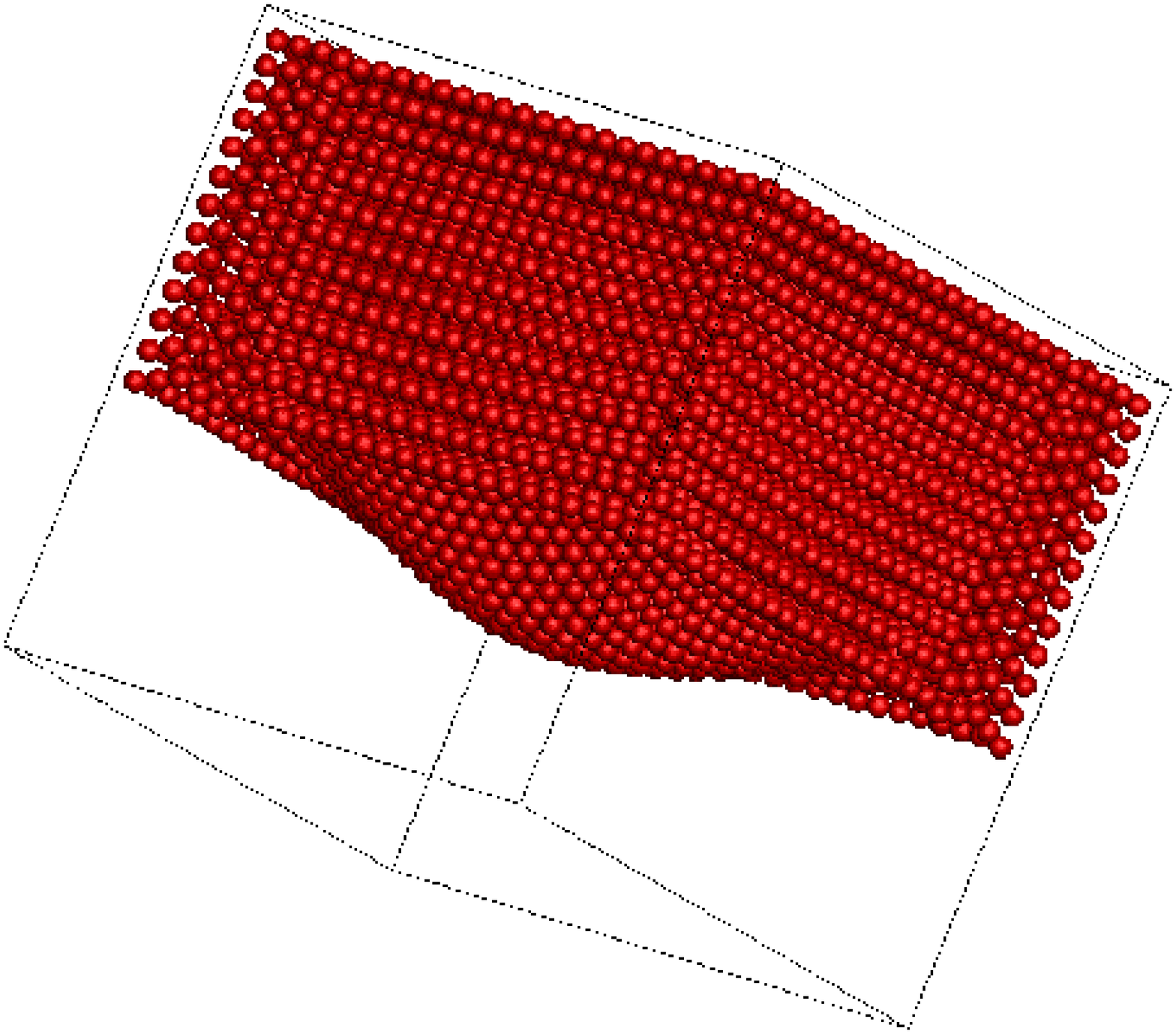}
\hspace{2mm}
\includegraphics[angle=  0,width=0.47\textwidth]{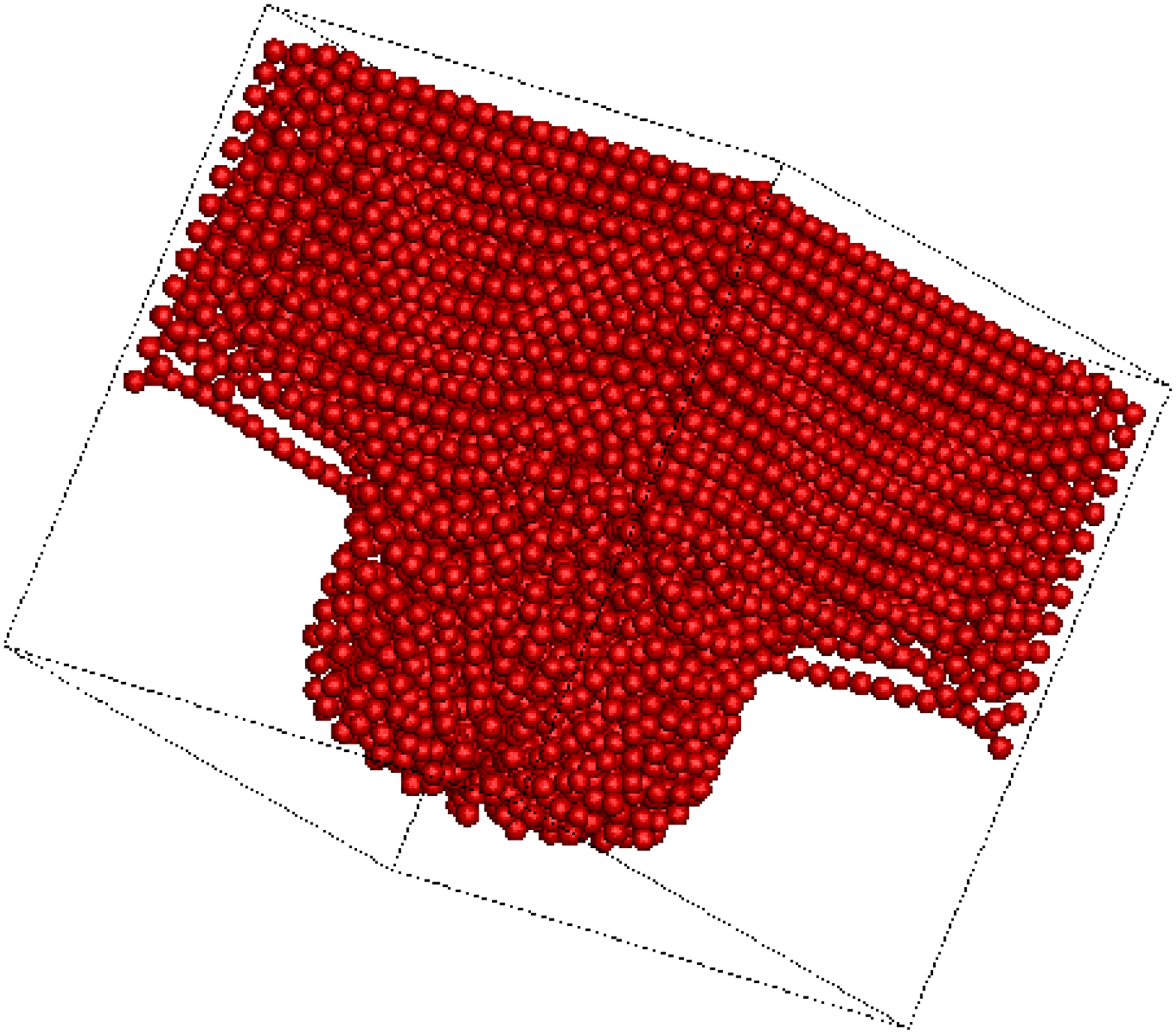}
\\
(a) \hspace{52mm} (b)
\\
\includegraphics[angle=  0,width=0.47\textwidth]{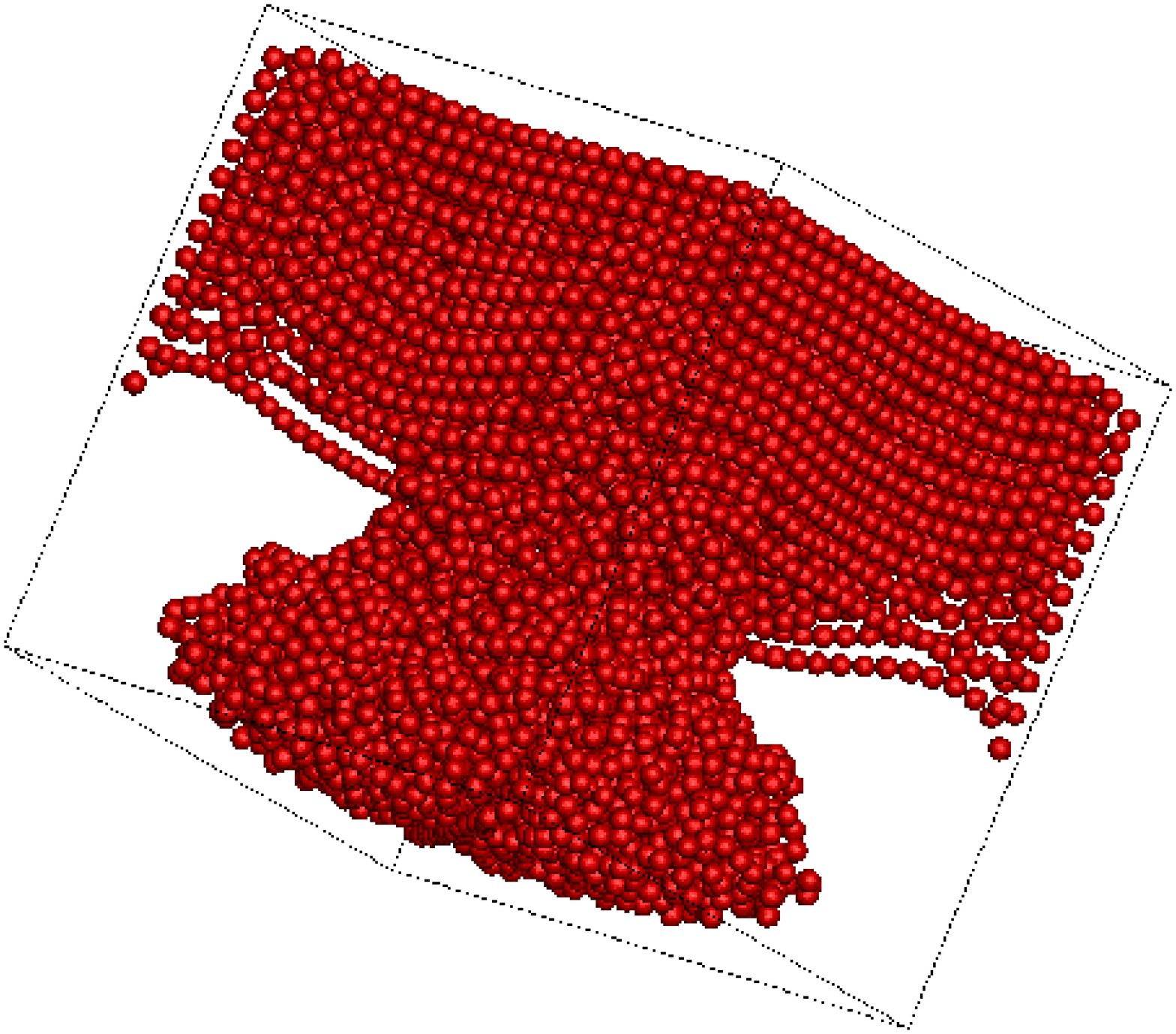}
\hspace{2mm}
\includegraphics[angle=  0,width=0.47\textwidth]{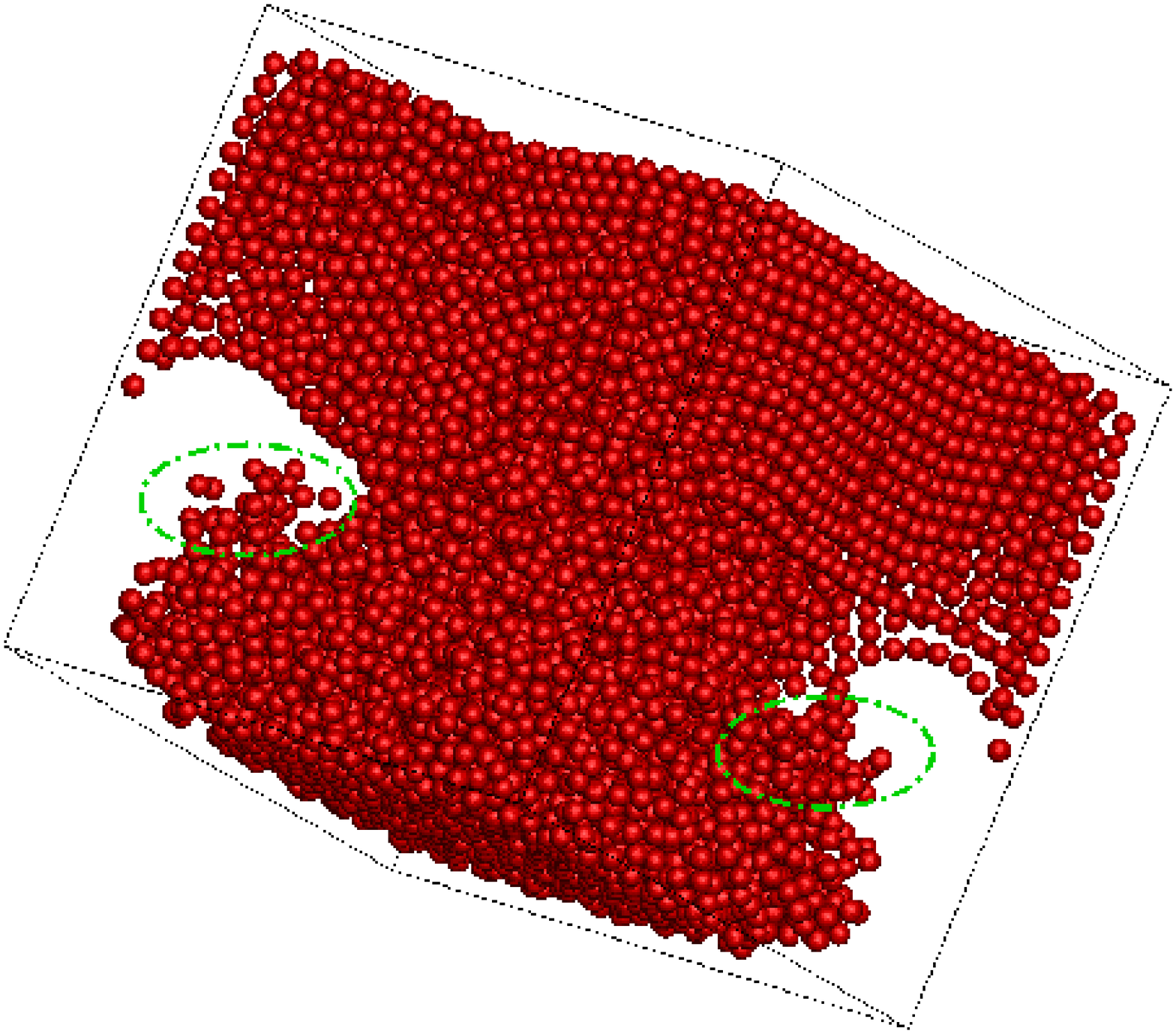}
\\
(c) \hspace{52mm} (d)
\caption{Positions of the 8125 particles at time 
(a) $t=2.5 s$, (b) $t=5.0 s$, (c) $t=7.5 s$, (d) $t=10.0 s$.}
\label{8125zao}
\end{figure}

\begin{figure}[!ht]
\centering
\includegraphics[angle=  90,width=0.45\textwidth]{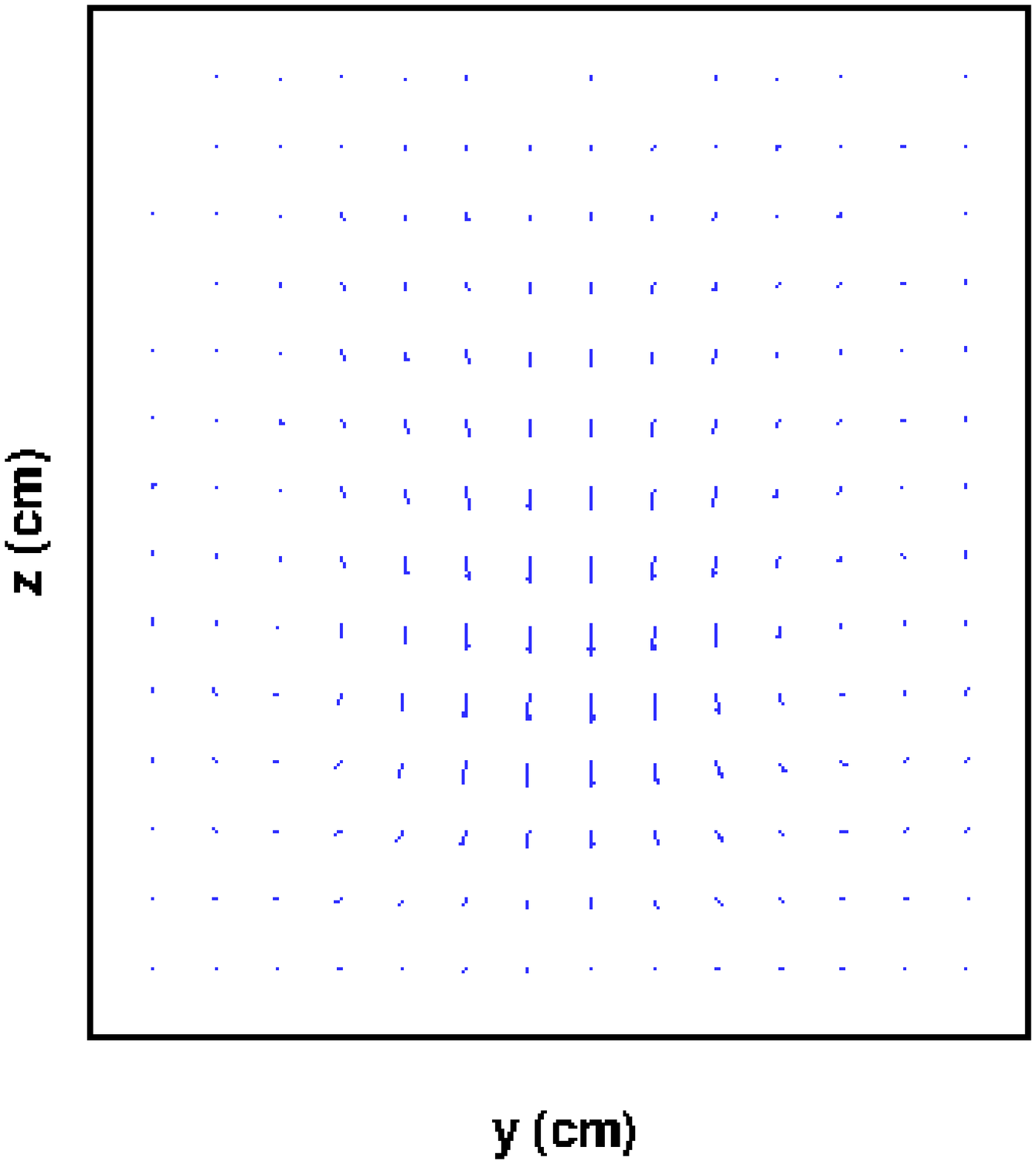}
\hspace{2mm}
\includegraphics[angle=  90,width=0.45\textwidth]{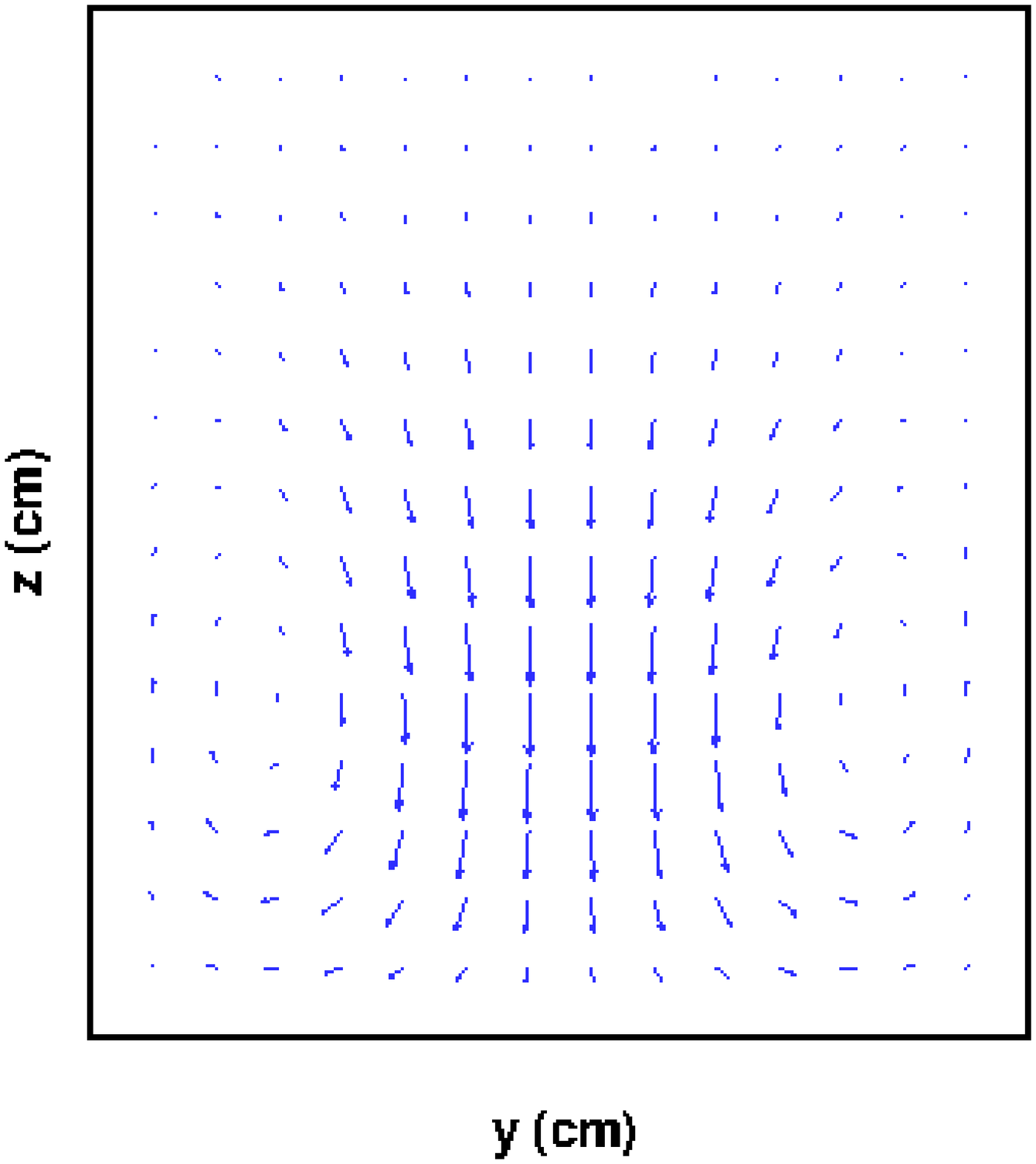}
\\
(a) \hspace{52mm} (b)
\\
\includegraphics[angle=  90,width=0.45\textwidth]{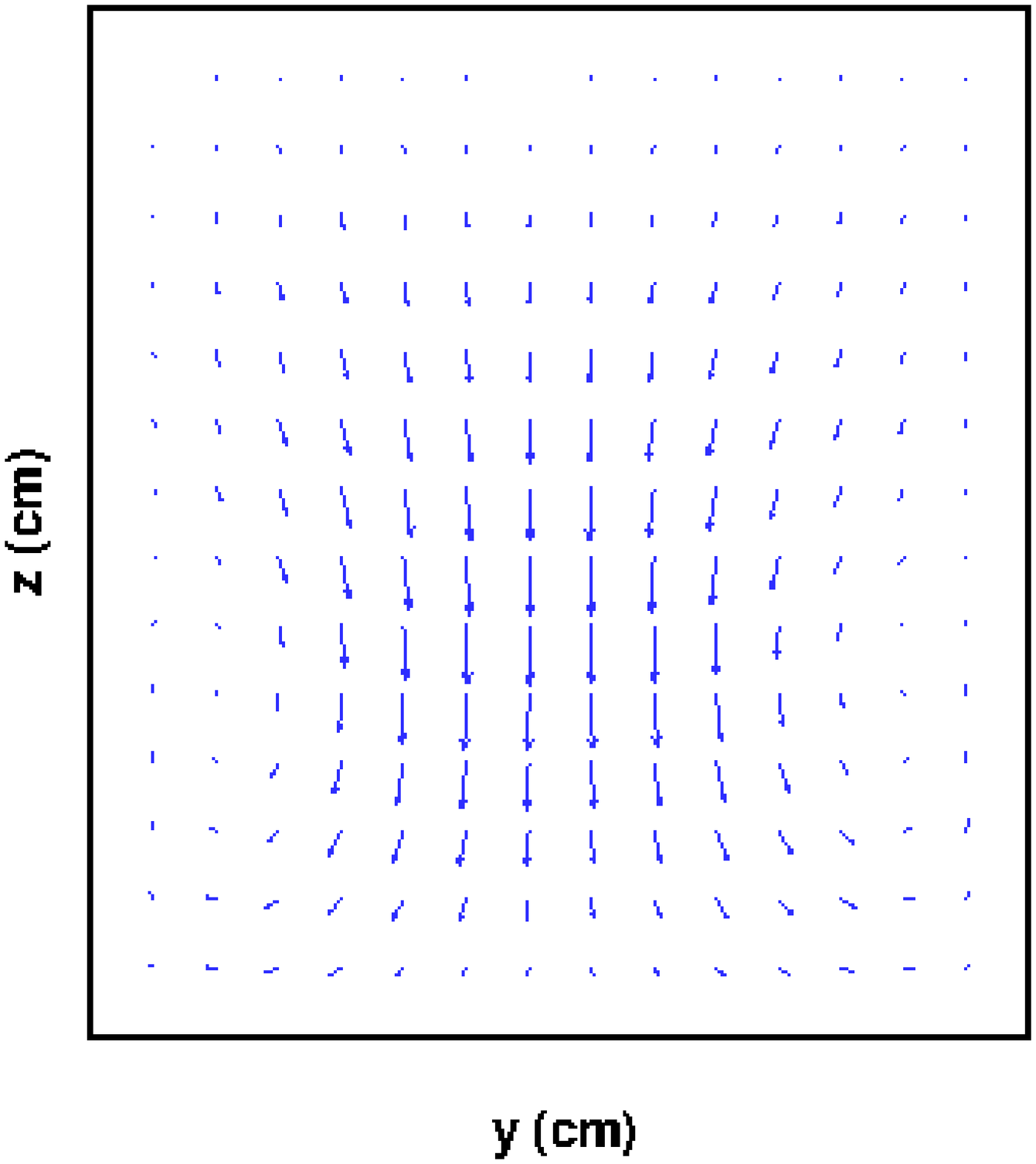}
\hspace{2mm}
\includegraphics[angle=  90,width=0.45\textwidth]{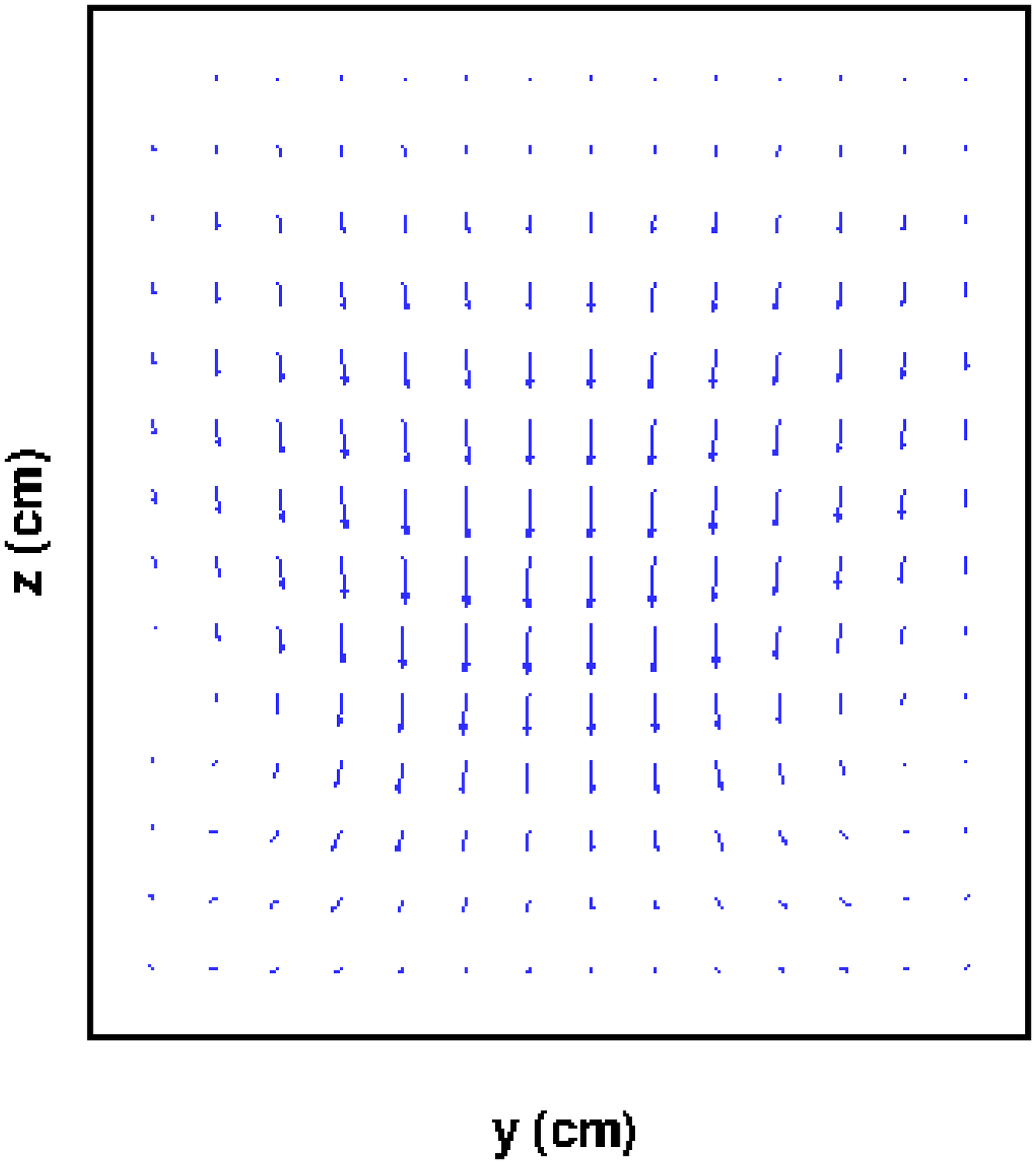}
\\
(c) \hspace{52mm} (d)
\caption{Instantaneous fluid velocity distribution on the mid-length slice at time 
(a) $t=2.5 s$, (b) $t=5.0 s$, (c) $t=7.5 s$, (d) $t=10.0 s$.}
\label{3dvelfluid}
\end{figure}

In the initial stages of sedimentation, an overall falling of the particle agglomeration can be observed as shown in 
Figure~\ref{8125zao} (a) and (b). Due to the fact that the initial porosity is low, the whole body at this 
stage can be regarded as a plug flow creeping in a channel. The distance between the highest particles and the top wall 
increases gradually and the particle distribution close to the walls does not change significantly. However, instead of settling 
uniformly, the difference of particle velocity inside and at the bottom of the body shows up shortly. This is because 
the particles close to the side walls are hindered by the stagnated fluid. Consequently, the particles in the center region 
move faster and pour down to suck the fluid to 
fill up the forming gap. The hump grows fast until it reaches the cavity bottom. It can be seen that the changing histories 
of the fluid-particle interface are different in two- and three-dimensional simulations. In the two-dimensional case, the
updraft of the fluid takes place mainly in the center. However, the three-dimensional particle-constructed pestle is too 
strong to break as shown in Figure~\ref{8125zao} (b) and the fluid is pushed away to take a devious route (explained later in 
Figure~\ref{3dvelfluid}). This observation is in line with the three-dimensional results reported by 
Robinson et al.~\cite{Robinson2014121} using Smoothed Particle Hydrodynamics (SPH)-DEM simulation. 
The discrepancy between two- and three-dimensional results is obviously due to the drawback of the two-dimensional assumption.

\begin{figure}[!ht]
\centering
\includegraphics[angle=  0,width=0.47\textwidth]{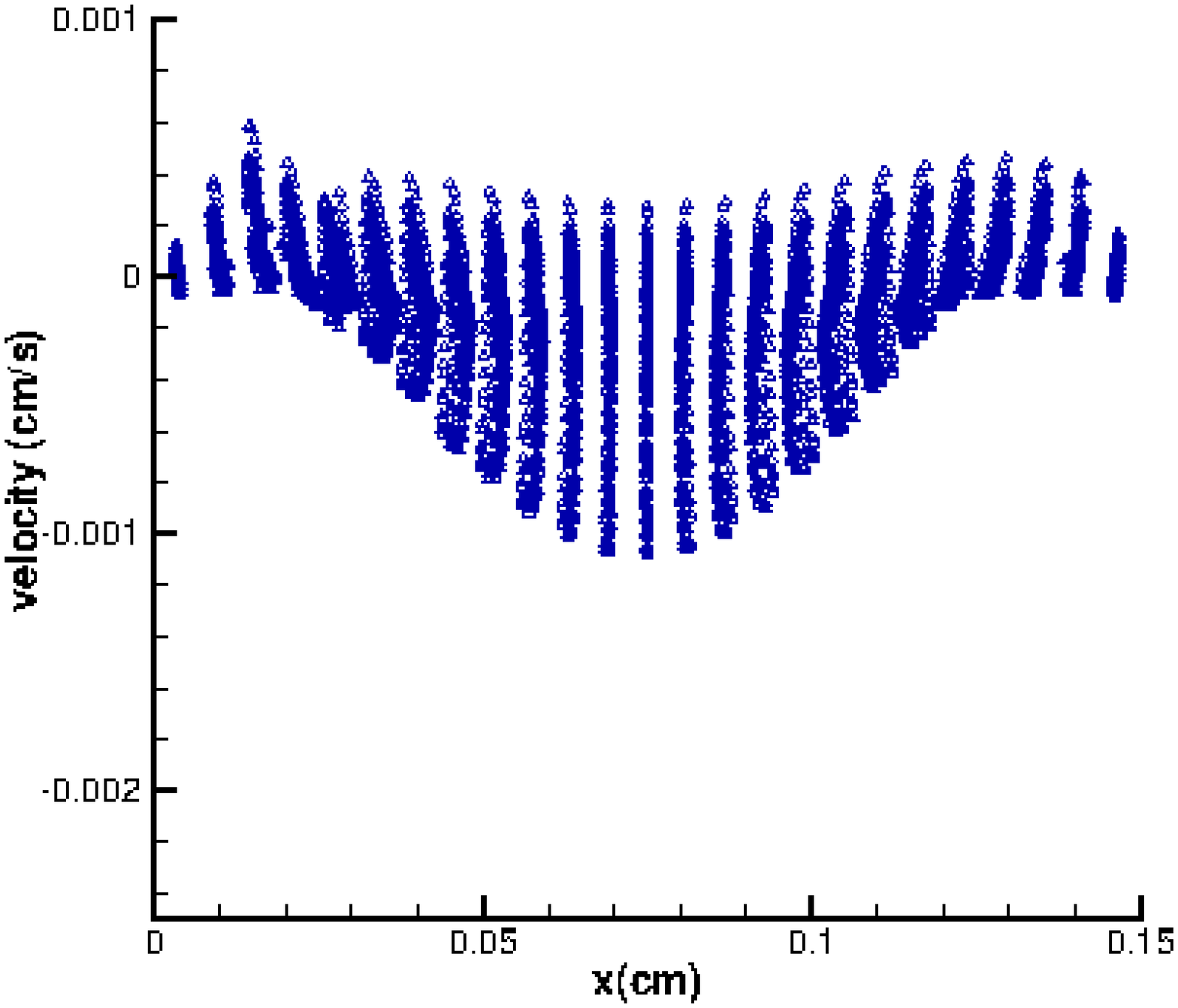}
\hspace{2mm}
\includegraphics[angle=  0,width=0.47\textwidth]{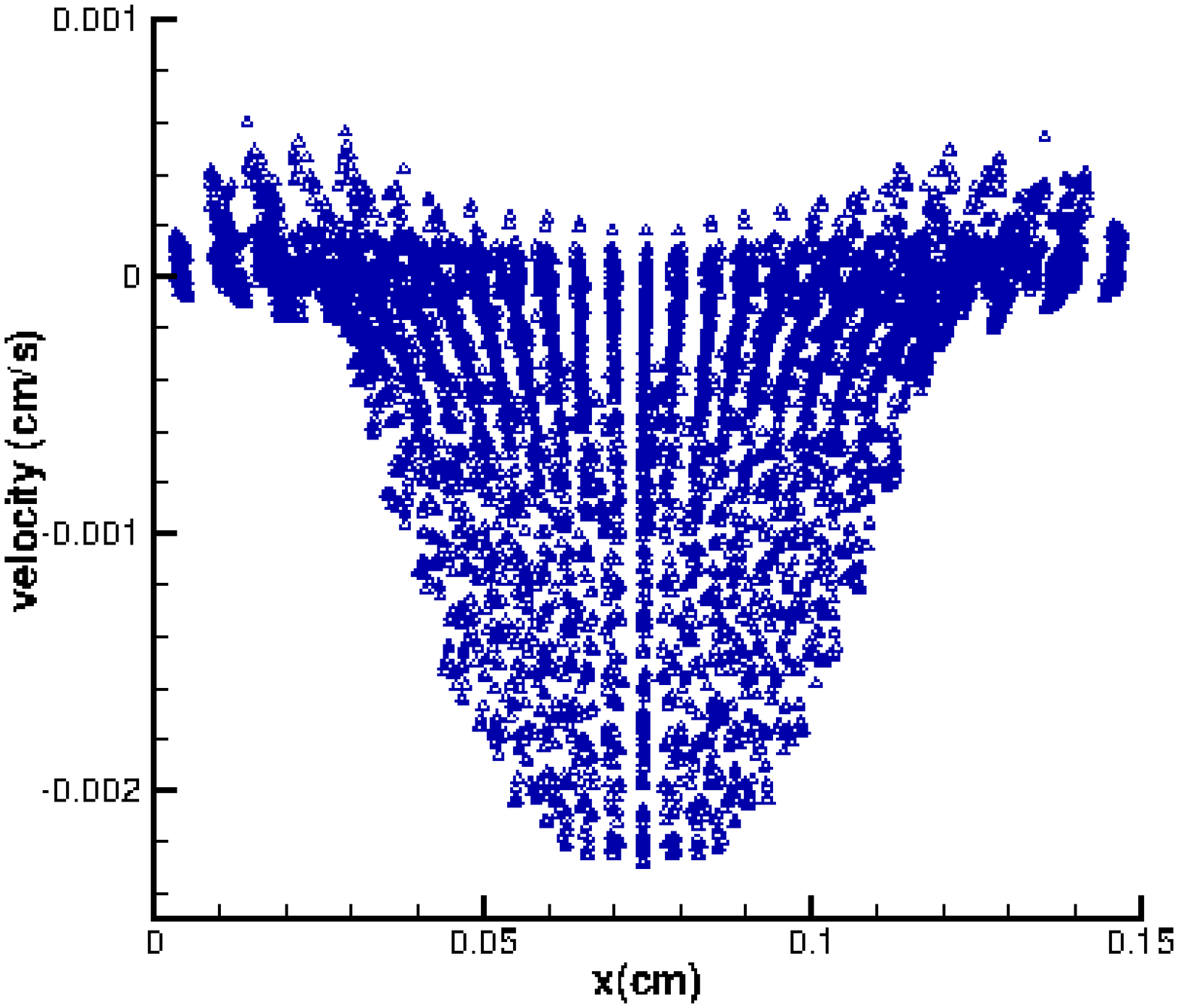}
\\
(a) \hspace{52mm} (b)
\\
\includegraphics[angle=  0,width=0.47\textwidth]{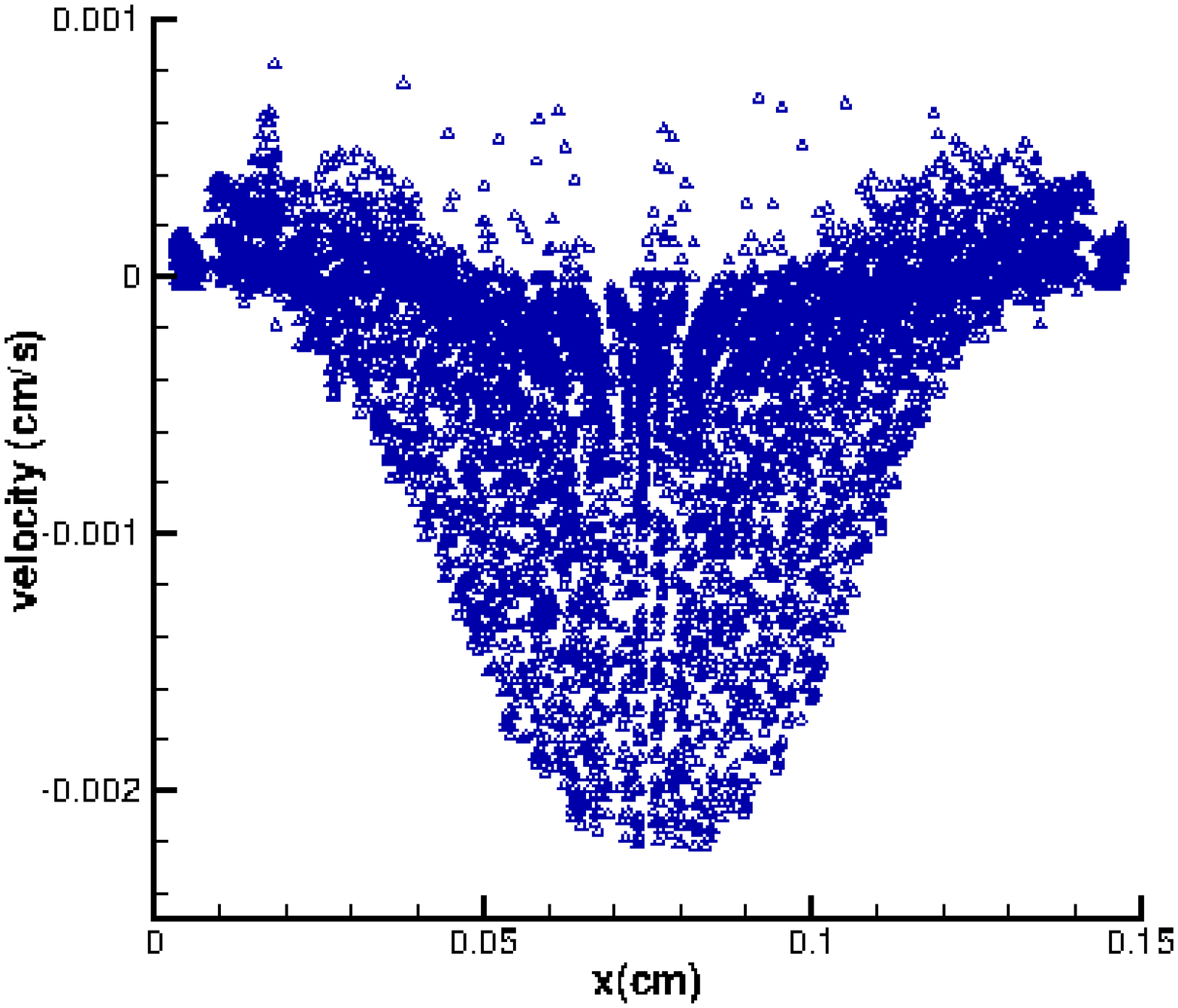}
\hspace{2mm}
\includegraphics[angle=  0,width=0.47\textwidth]{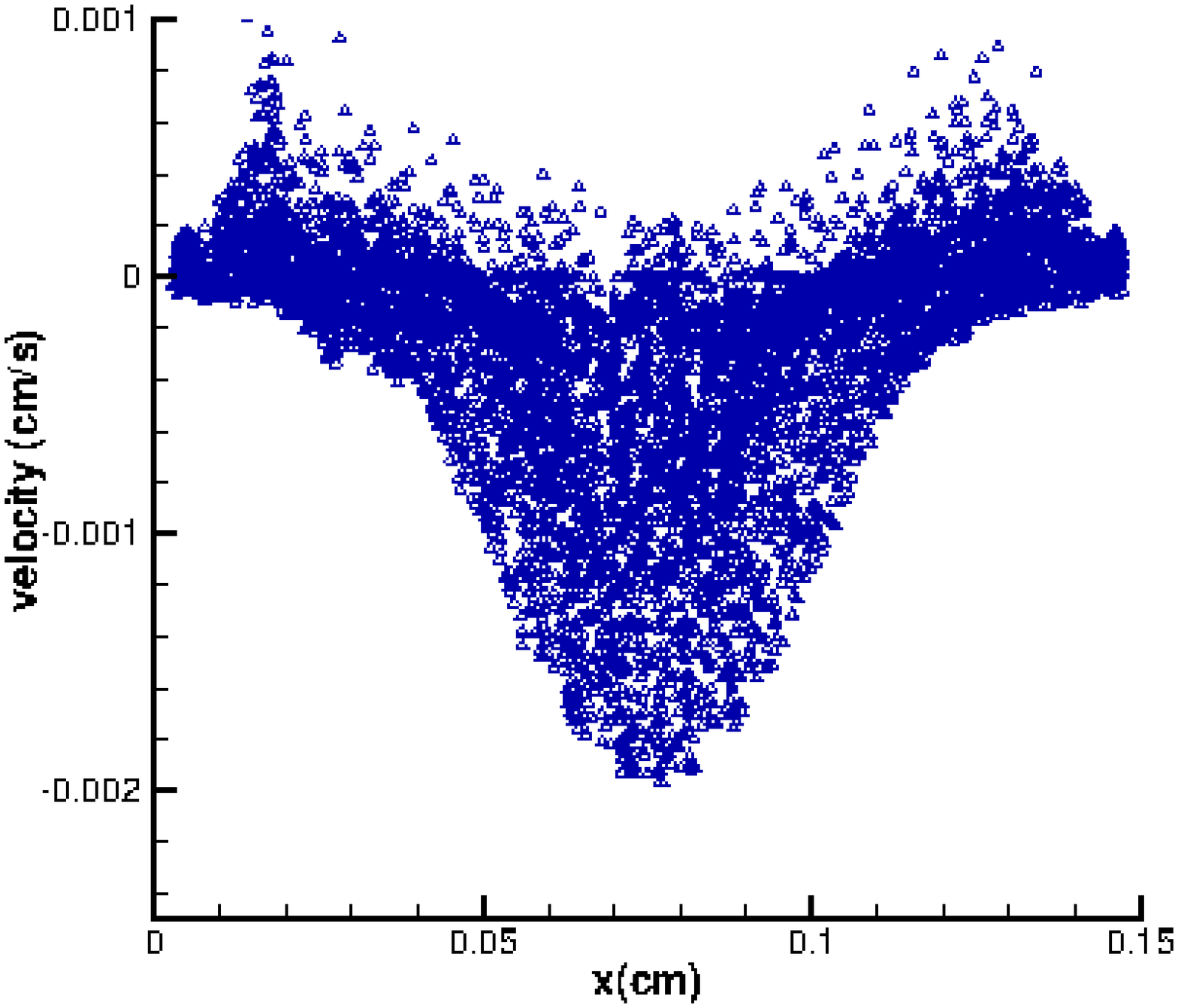}
\\
(c) \hspace{52mm} (d)
\caption{Particle deposition velocity along the $x-$direction at time 
(a) $t=2.5 s$, (b) $t=5.0 s$, (c) $t=7.5 s$, (d) $t=10.0 s$.}
\label{3dvelocity}
\end{figure}

The following three-dimensional deposition processes show nearly opposite trend comparing with the two-dimensional case as 
shown in Figure~\ref{8125zao} (c) and (d). In the two-dimensional case, a fluid hole is formed in the 
lower half of the cavity which is hugged by two particle arms, this typical phenomenon has been reported in several 
 studies~\cite{glowinski1999distributed,Feng2004602,Feng200520,ZHANGCAF2014}.
However, in the three-dimensional case, it is more like a fluid hoop surrounding the particle pestle. 
The head of the pestle spreads out when it impacts on the bottom. The behavior is not difficult to understand 
because the successive falling particles keep moving downward and thus pushing on the head. At this time, 
the underriding of the particles becomes the dominating force in the system and most of 
the particles distribute in this center region. 

\begin{figure}[!ht]
\centering
\includegraphics[angle=  0,width=0.45\textwidth]{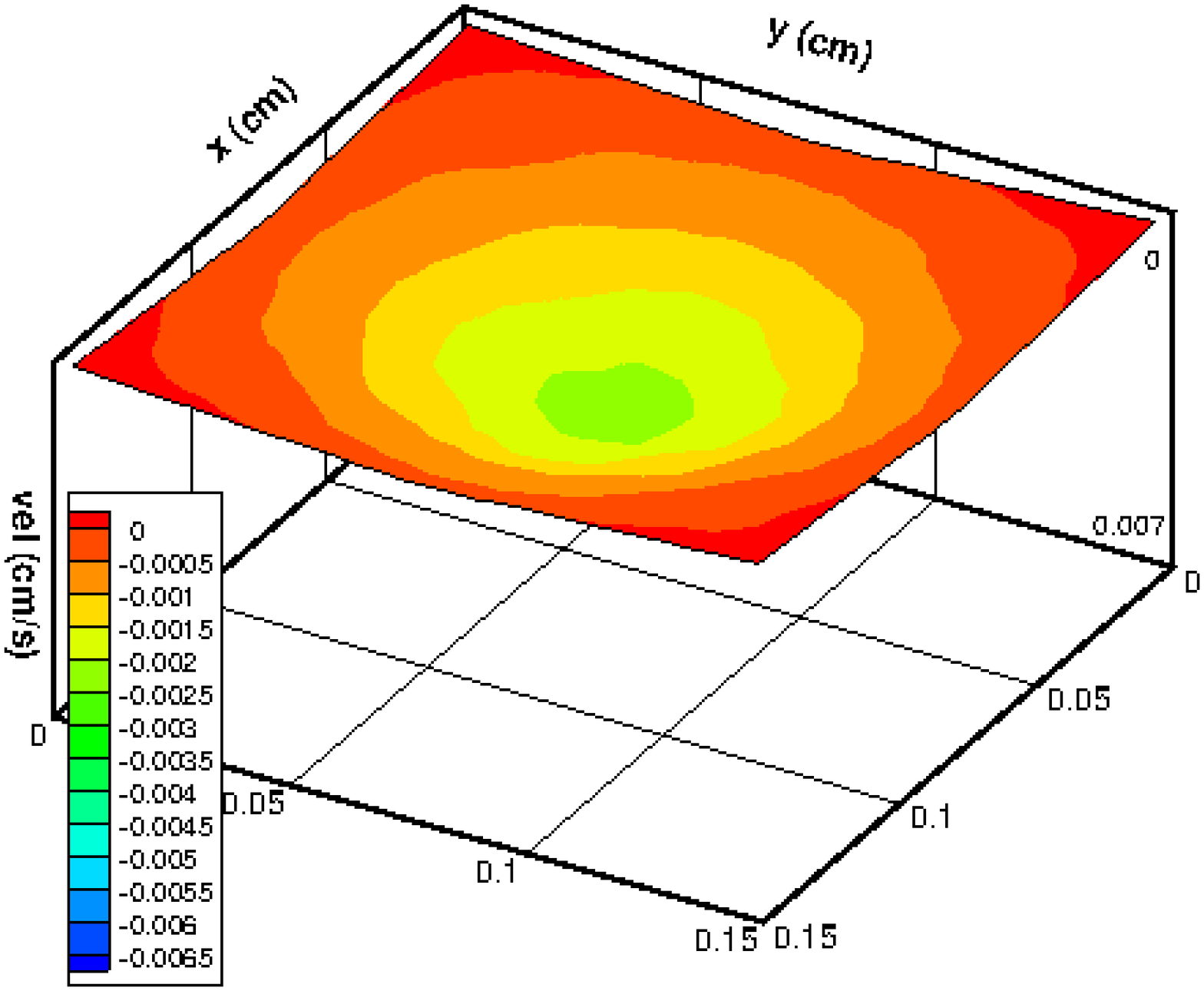}
\hspace{2mm}
\includegraphics[angle=  0,width=0.45\textwidth]{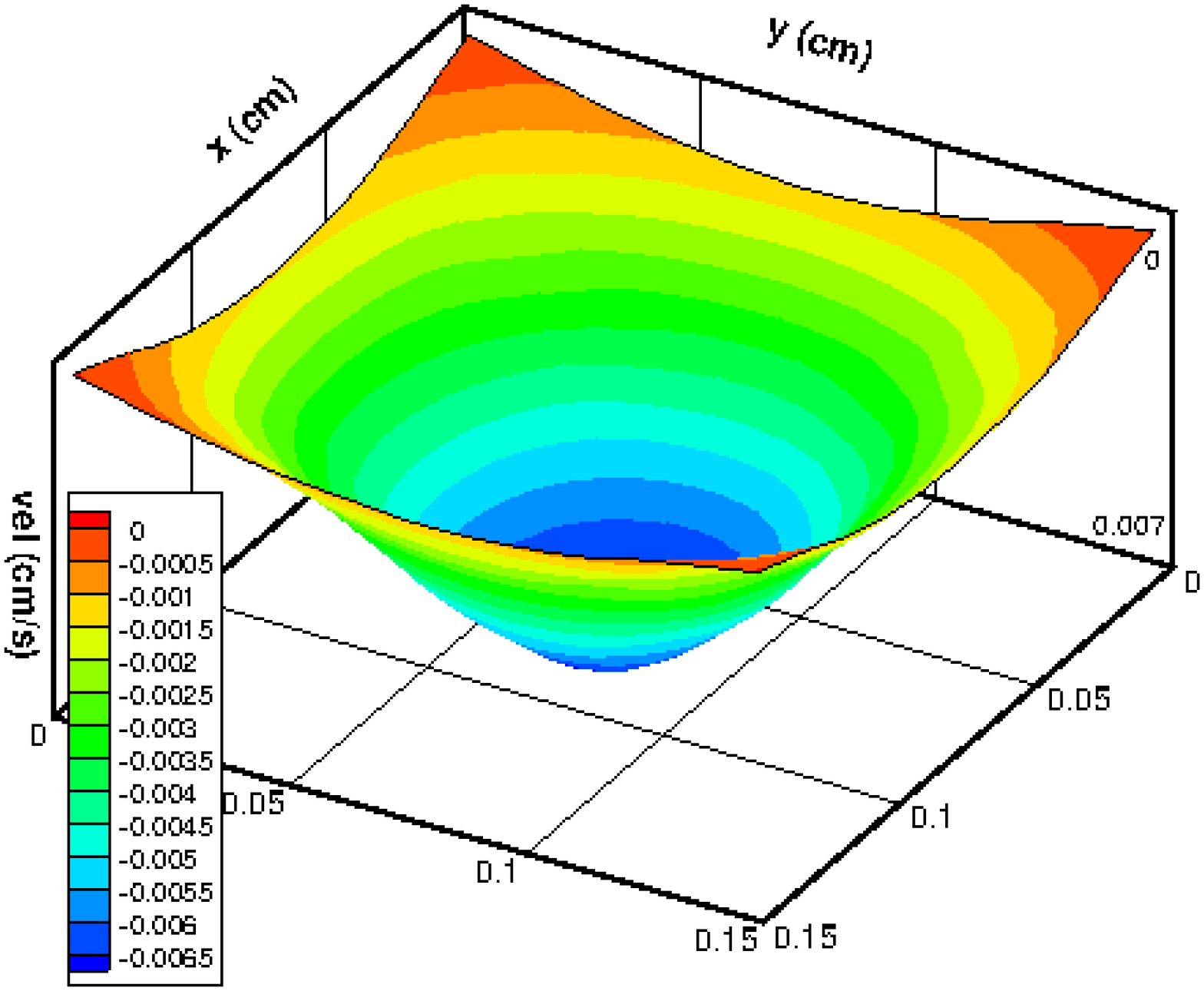}
\\
(a) \hspace{52mm} (b)
\\
\includegraphics[angle=  0,width=0.45\textwidth]{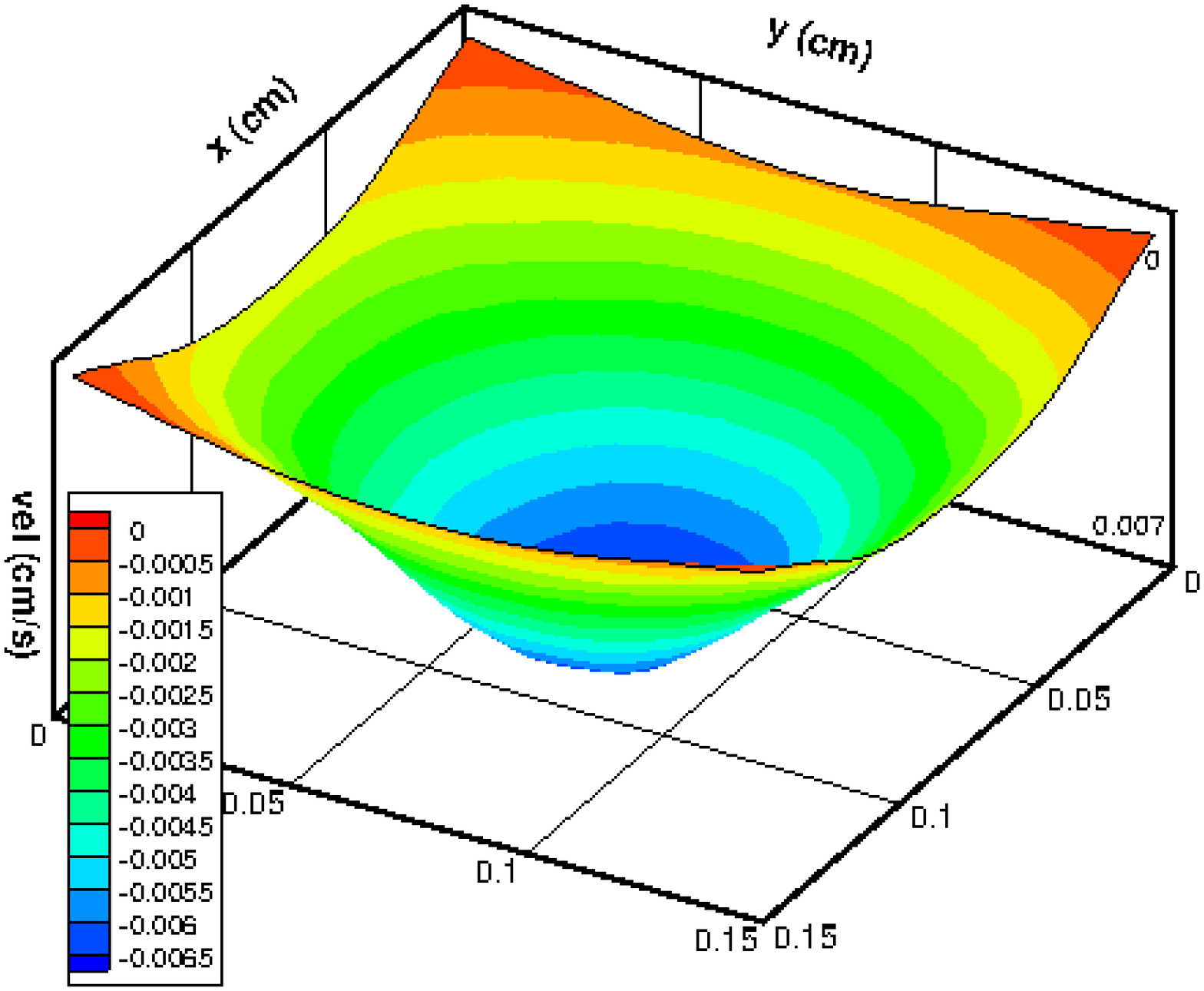}
\hspace{2mm}
\includegraphics[angle=  0,width=0.45\textwidth]{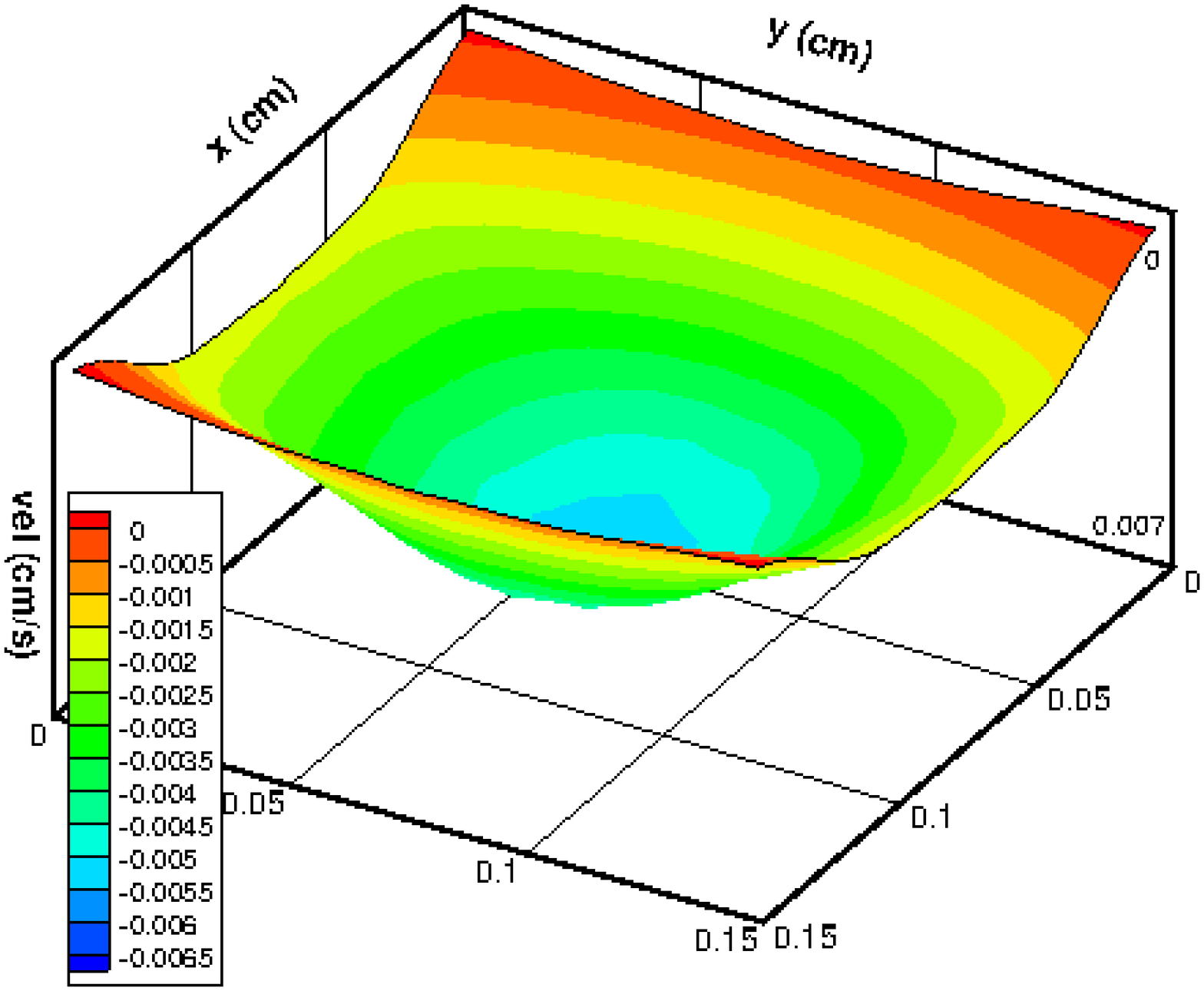}
\\
(c) \hspace{52mm} (d)
\caption{Overall distribution of the particle deposition velocity at time 
(a) $t=2.5 s$, (b) $t=5.0 s$, (c) $t=7.5 s$, (d) $t=10.0 s$.}
\label{3dveldis}
\end{figure}

Since the particle deposition velocities are very important for the efficiency of the final deposition
and may lead to a non-uniform distribution on the bottom. Figure~\ref{3dvelocity} displays the distributions 
of particle velocity along the $x-$direction before $10.0 s$ where large discrepancy can be observed. It is shown 
that most of the velocities have negative signs and the larger deposition velocities concentrate in the center region. 
This finding is in line with the particle distribution patterns. Moreover, the magnitude of the deposition velocity 
increases with time until the particles impact on the bottom. It is also clearly seen that the majority of the velocities 
in the regions close to the side walls are positive due to the fact that the sucked fluid 
pushes the high particles up when the center particles sink down. This interesting FSI phenomenon can be clearly observed 
in Figure~\ref{3dvelfluid} where the instantaneous fluid velocity corresponding to Figure~\ref{3dvelocity} 
is given. It is shown that the initial stagnant fluid is disturbed by the particle motion and follows the trend of the 
solid particles. Two vortexes (hoop in three-dimensional geometry) are formed in the lower corners of the cavity and the 
fluid velocity near the side wall is upward. The vortexes are strong when the particle deposition velocities are large. 
As shown in Figure~\ref{3dvelocity} (c) and (d), the particle deposition velocities begin to decrease after the particles 
reach the bottom, meanwhile the number of particles with positive velocities increases. These particles are risen by the 
vortexes and against the falling particles as shown in Figure~\ref{8125zao} (d) highlighted by green ellipse. An overall 
distribution of the particle deposition velocities is given in Figure~\ref{3dveldis} in terms of mean values.
Here, the whole bottom domain is divided by $30\times30$ squares and then the particles are mapped into the square 
that the particle center lies. The square holds the deposition velocity that mapped in it. If more than one particle 
is mapped into the same square, the arithmetic mean value will be employed. As shown, the mean velocities present a 
generally symmetrical distribution. The particles near the corners deposit significantly slower than the center as 
results of the fluid viscosity. The highly symmetrical distribution is broken when the particle contact with the bottom. 
However, a constant symmetrical distribution may not be expected due to the stochastic nature of the solid particles. 
From $t=10.0 s$, the collisions between the particles and particles/walls become the dominating force in the lower half 
of the cavity. The pestle slumps like an inverted cone and fills the cavity bottom.

\begin{figure}[!ht]
\centering
\includegraphics[angle=  0,width=0.47\textwidth]{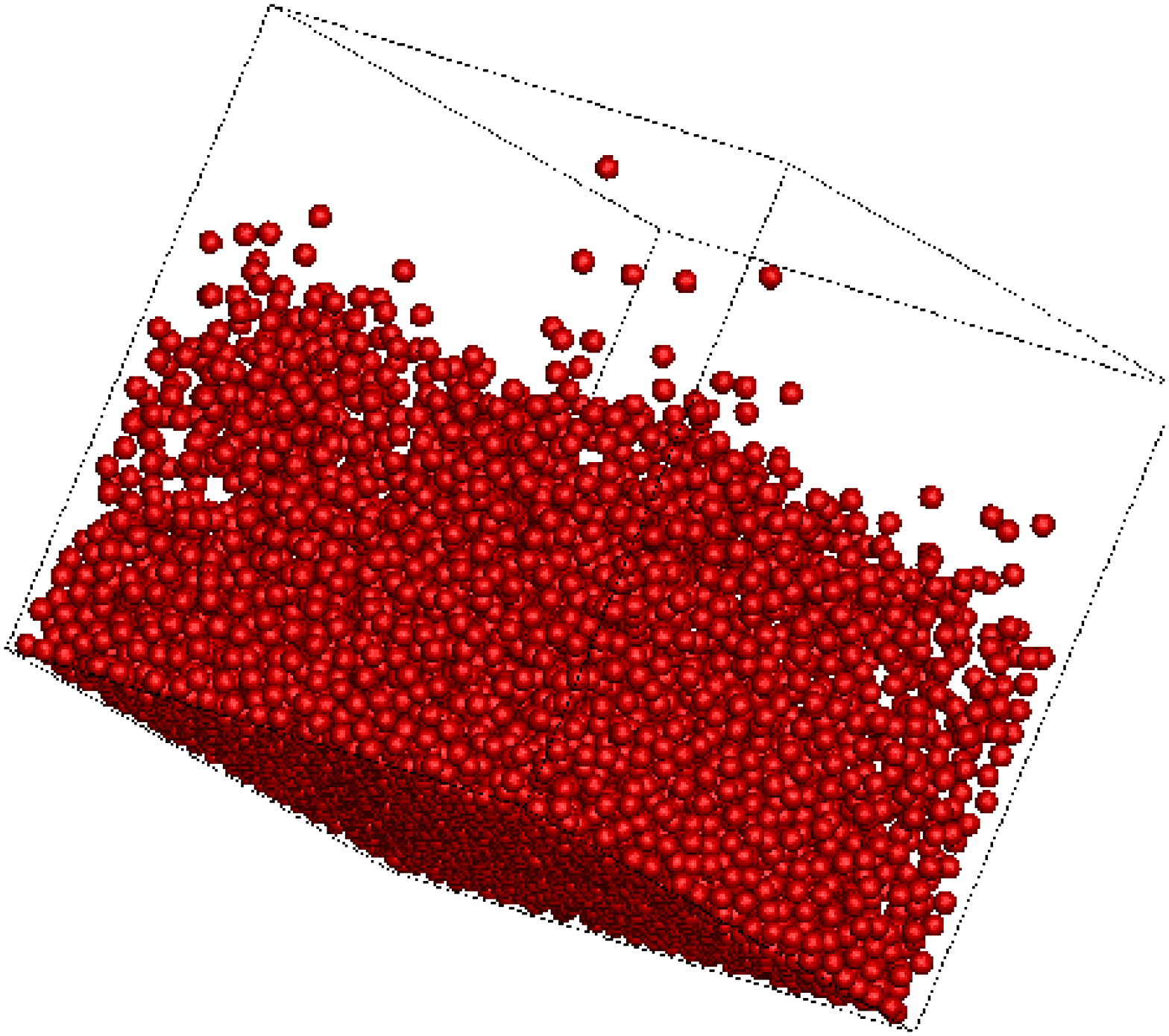}
\hspace{2mm}
\includegraphics[angle=  0,width=0.47\textwidth]{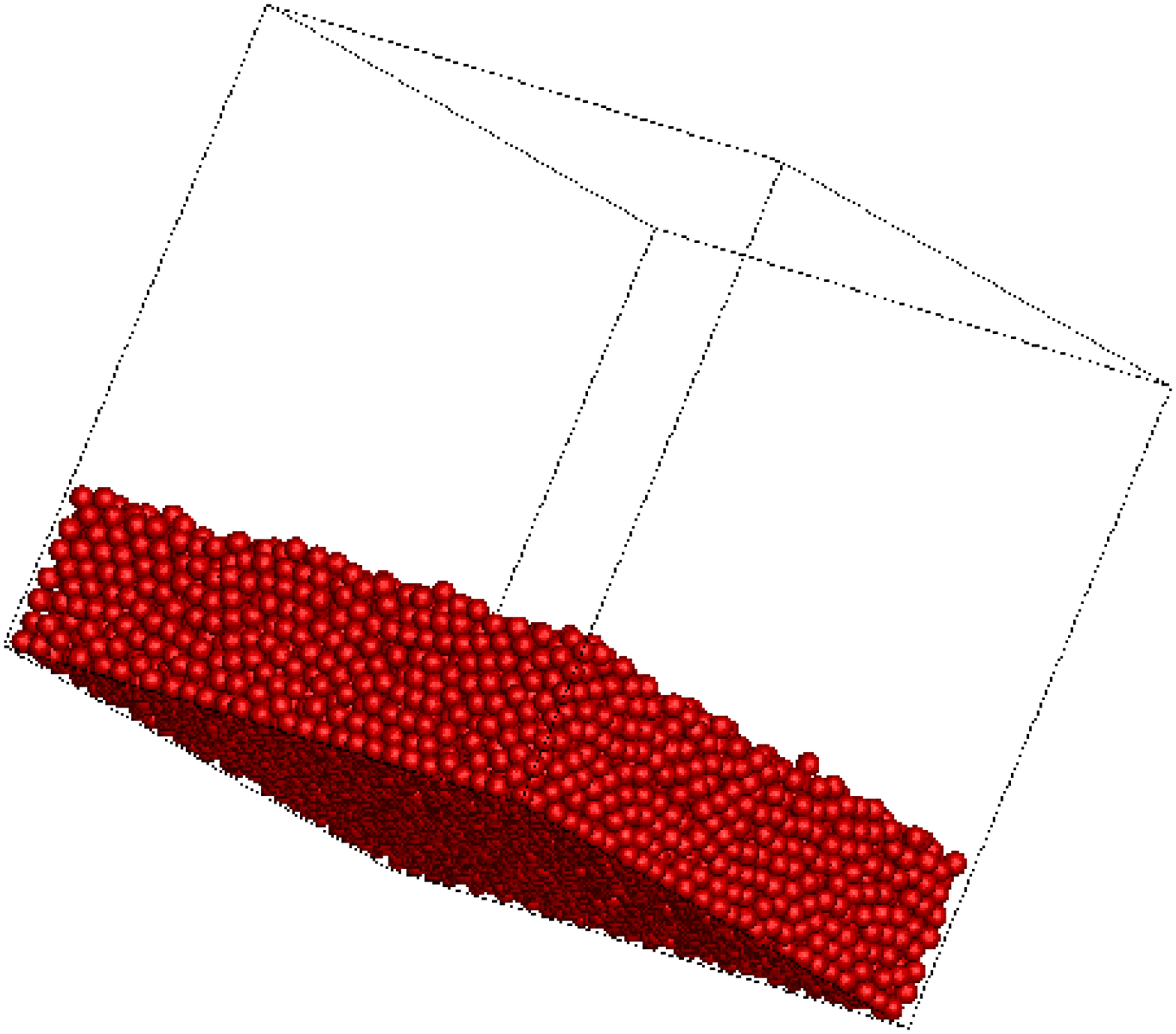}
\\
(a) \hspace{52mm} (b)
\caption{(a) Positions of the 8125 particles at time $t=50.0 s$ and (b) the final distribution.}
\label{5k3wan}
\end{figure}

Figure~\ref{5k3wan} (a) and (b) display the later stages of the depositing process. In Figure~\ref{5k3wan} (a),
the initially orderly arranged particles are totally disorganized and settle on the cavity bottom length by length.

\subsubsection{Effect of the initial porosity}

\begin{table}[!ht]
\centering
\begin{tabular}{cccc}
  &  \\ 
\hline
Particle number &  Initial porosity  & Solid fraction & Initial distribution    \\
          
\hline
$8125$    & $0.719$ & $0.15$  & $25\times25\times13$  \\   
$5200$    & $0.813$ & $0.10$  & $20\times20\times13$   \\ 
$2925$    & $0.888$ & $0.056$  & $15\times15\times13$   \\ 
\hline
\end{tabular}
 \caption{The settling velocities at different particle size.}
\label{lporosity}
\end{table}

It has been well known that the porosity can play an important role in the sedimentation of multi particles. 
Here, different numbers of particles were positioned in the same region as previous subsection. 
In other words, the particles would deposit with different initial porosity. 
The physical properties of the fluid and particles can be found in Table~\ref{Stokes}. 
The minimum particle height was monitored to characterize the sedimentation efficiency.
The parameters relevant to these simulations are listed in Table.~\ref{lporosity}.

\begin{figure}[!h]
\centering
\includegraphics[width=0.5\textwidth]{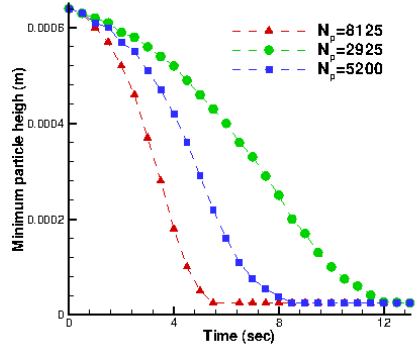}
\vskip-0.2cm
\caption{Minimum particle height versus time at different initial porosity.} \label{droptime}
\end{figure}

Figure~\ref{droptime} shows the minimum particle height versus time with different initial porosity. 
It can be seen that the sedimentation efficiency increases with the decrease of the initial porosity 
even identical particles were used, this finding is consistent with the analytical results from 
Robinson et al.~\cite{Robinson2014121}. Moreover, a significant deceleration of settling velocity 
can be observed when the lowest particles approach to the cavity bottom, this phenomenon has also been
reported in ~\cite{Feng200520} and ~\cite{ZHANGCAF2014}  . 

\subsubsection{Effect of the particle number on the total computational cost}

\begin{figure}[!h]
\centering
\includegraphics[width=0.5\textwidth]{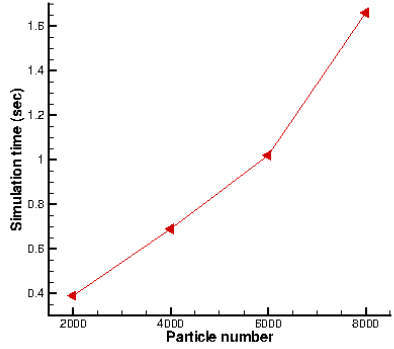}
\vskip-0.2cm
\caption{Particle number versus the computational time in one time step.} \label{pnvstime}
\end{figure}

At last, for the sake of examining the effect of the particle number (the number of the Lagrangian point in conventional IBM)
on the total
computational cost, several simulations were carried out with different particle number. As shown in Figure~\ref{pnvstime},
the total computational cost increases almost linearly with the particle number and the slope is even larger when the particle 
number increases from 6000 to 8000. It is worthwhile mentioning that Figure~\ref{pnvstime} was obtained when there are no 
particle collisions in the system. We also tested the computing time of each part of the solver in above 8125 particle 
simulation at time $t=30.0 s$, we found that the calculation of the fluid-particle interaction force spends about $84.4 \%$ 
of total simulation time in one time step
and the total particle collision number is 6610. Therefore, we come to a conclusion that the total computational cost 
can be significantly reduced by decreasing the number of the Lagrangian point. Comparing with the conventional 
LBM-IBM-DEM~\cite{ZHANGCAF2014}, dozens of times (divided by $N_{LP}$) speedup can be expected in two-dimensional simulation 
and hundreds of times in three-dimensional simulation under the same particle and mesh number.
However, it is worthwhile mentioning that this conclusion is reached only from a computational efficiency point of view.
For a certain problem with large range of sizes of particles, a hybrid IBM-PIBM may be needed to achieve high performance 
calculation which will be discussed in next subsection.

Overall, the main findings of the two- and three-dimensional simulations are summarized as follows:
The patterns observed in the two-dimensional simulation are close to the results provided in former 
references~\cite{glowinski1999distributed,Feng2004602,Feng200520,ZHANGCAF2014}. However, the three-dimensional 
results show large discrepancy with the two-dimensional results which is most probably due to the two-dimensional 
assumption. Imaging a case in a enclosed container like a fluidization bed, the fluid is unpenetrable into a 
two-dimensional well-packed particle bed without breaking the compact structure, whereas penetration into a 
three-dimensional bed is somehow possible because the geometry is much more polyporous and complex.

\subsubsection{Hybrid IBM-PIBM modeling}

\begin{figure}[!ht]
\centering
\includegraphics[angle=  0,width=0.47\textwidth]{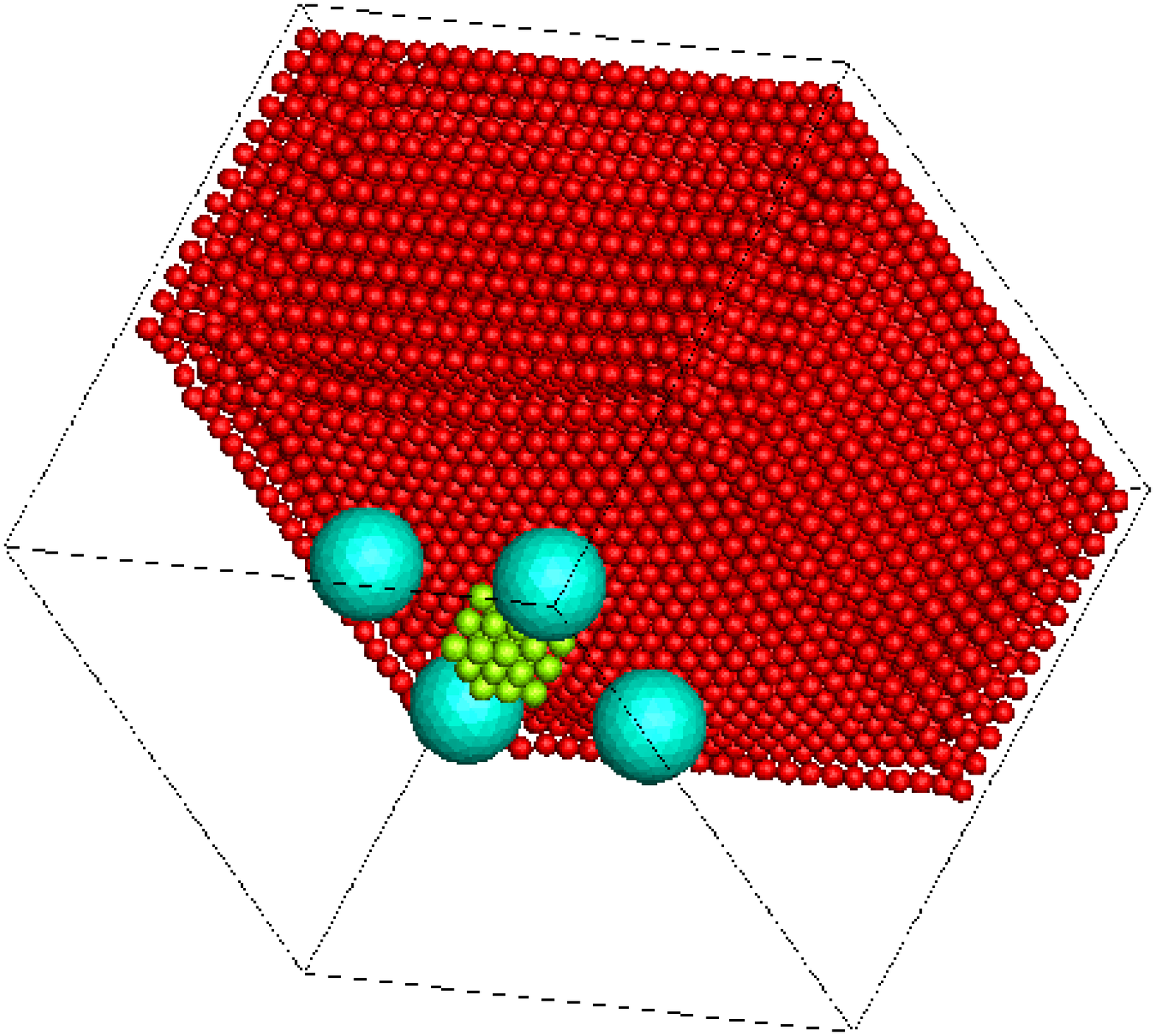}
\hspace{2mm}
\includegraphics[angle=  0,width=0.47\textwidth]{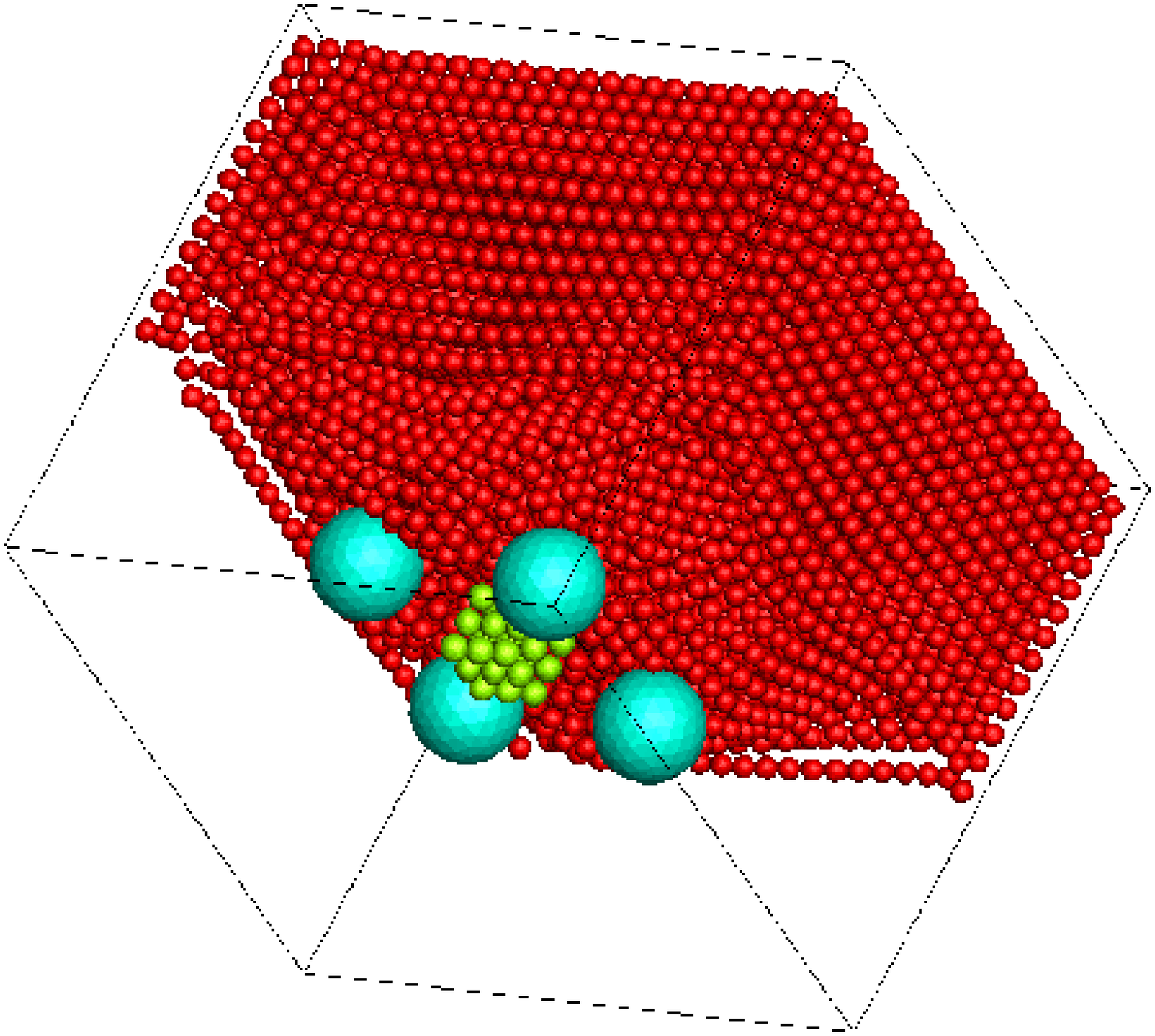}
\\
(a) \hspace{52mm} (b)
\\
\includegraphics[angle=  0,width=0.47\textwidth]{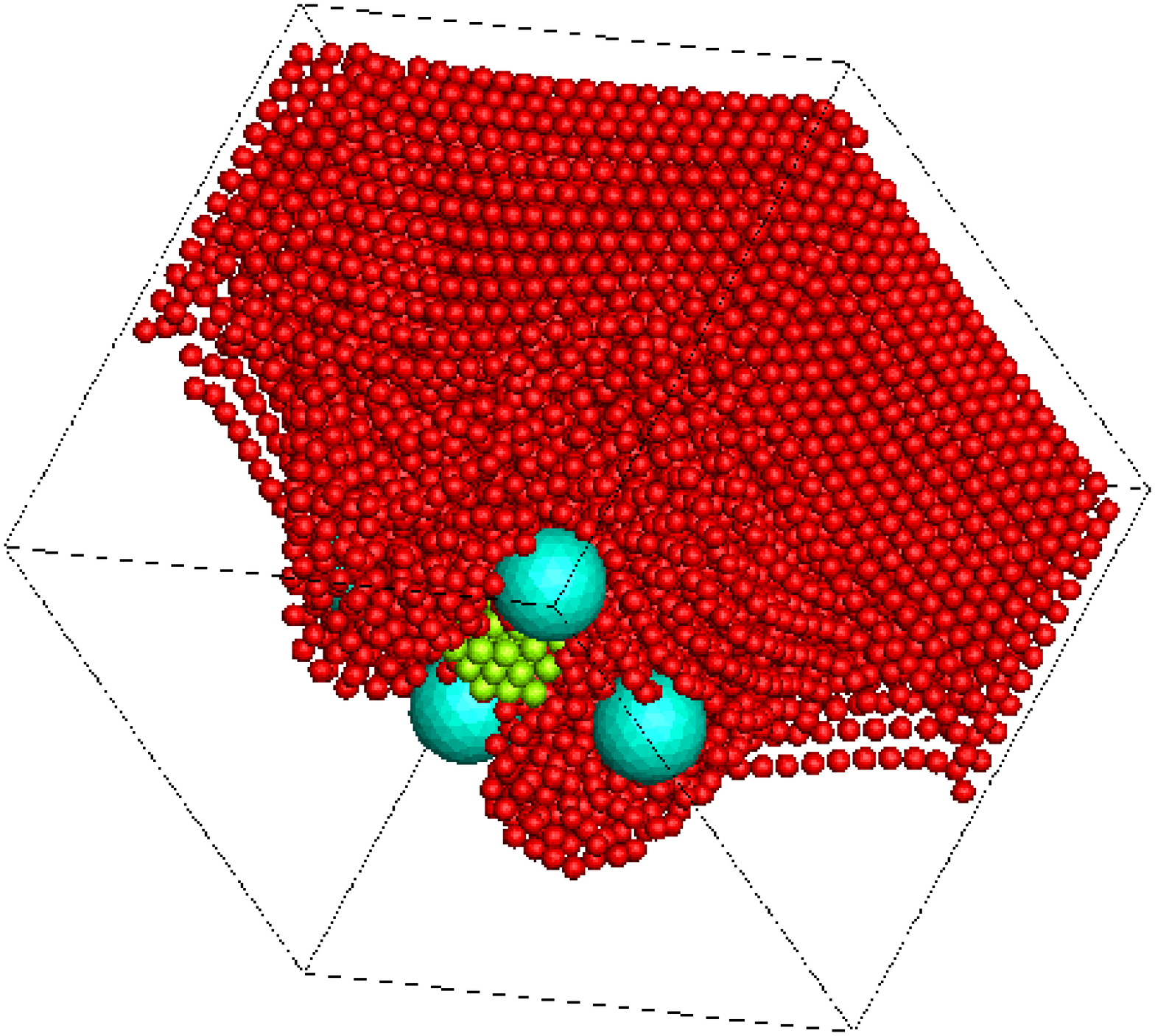}
\hspace{2mm}
\includegraphics[angle=  0,width=0.47\textwidth]{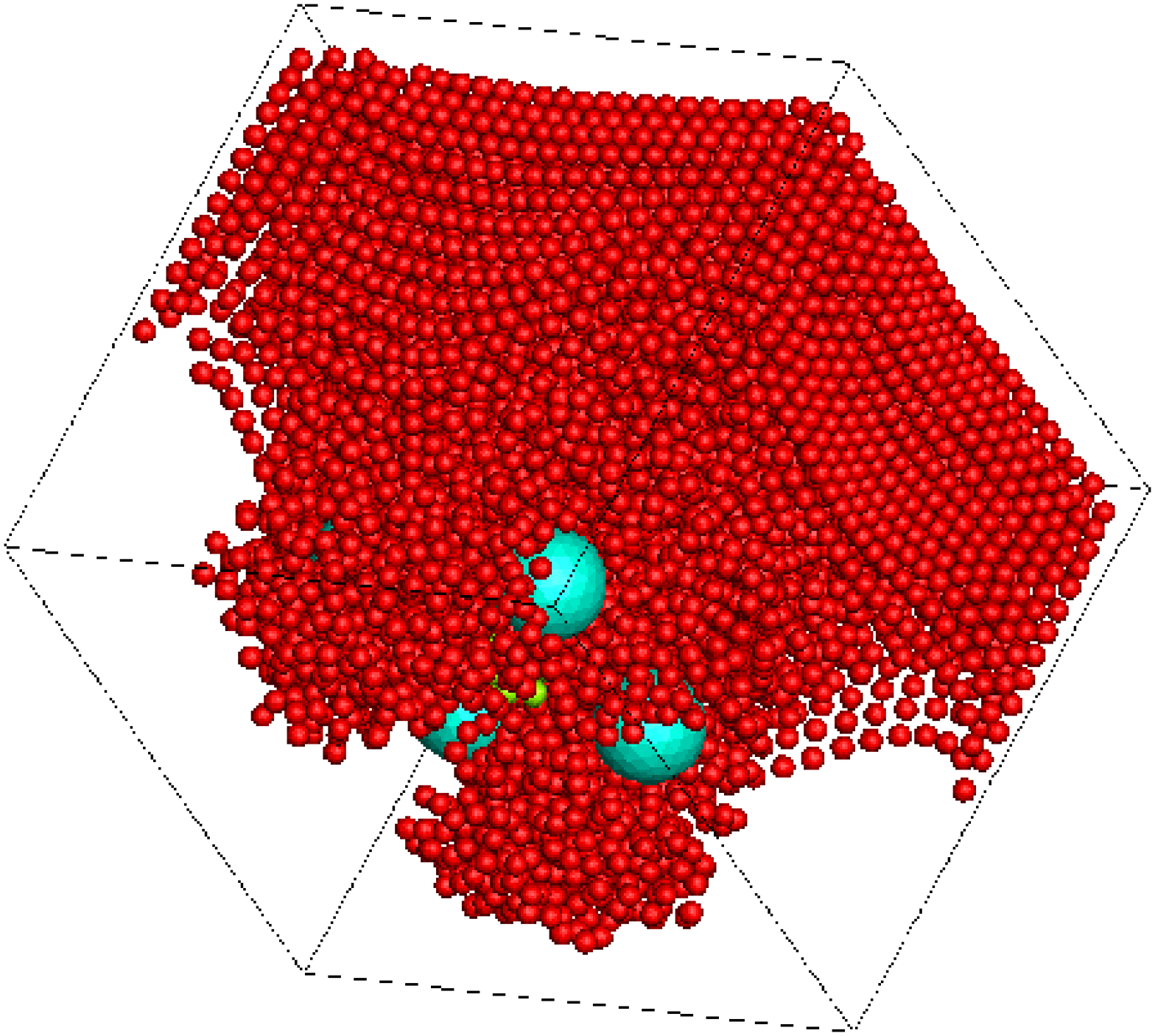}
\\
(c) \hspace{52mm} (d)
\caption{Positions of the 8125 particles with obstacles at time 
(a) $t=2.5.0 s$, (b) $t=5.0 s$, (c) $t=7.5 s$, (d) $t=10.0 s$.}
\label{8156mix}
\end{figure}

It is common to encounter a system containing various sizes of particles. A multiscale analysis is preferred when 
the size range is large. If all the particles are treated using the conventional IBM, the number of grids required to 
construct the finest particle would make the whole simulation too expensive. On the contrary, if all the coupling work 
is carried out based on the PIBM, the grid required to embody the largest particle would be too coarse to accurately 
reflect the fluid flow.  A more frequently encountered requirement is to build the complex boundaries or irregular elements
using IBM. In other words, a hybrid IBM-PIBM method is needed. Using a simply sample as shown in Figure~\ref{5k3wan}, the 
advantage of this mixed approach can be seen where five stationary obstacles are fixed below the particles (four large 
particles and one cube consisting of 27 small particles). The four large particles are established using the conventional 
IBM while the rest, including the 27 particle for the cube, are treated using PIBM. The criterion to choose different methods 
is the ratio between the particle size and the lattice spacing. In general, the grid size can be specified at 10 times the 
particle sizes in the Eulerian-Eulerian model~\cite{cloete2011fine} and about 5 times in the Eulerian-Lagrangian model based 
on NS-DEM~\cite{kafui2002discrete}. However, the results from current study indicate that the ratio can be 2 in LBM-PIBM-DEM 
though the optimal ratio is still in question.

\section{Concluding remarks}\label{conclusion}

A PIBM for simulating the particulate flow in fluid was presented. Compared with the conventional 
momentum exchange-based IBM, no artificial parameters are introduced and the implementation 
is simpler. The PIBM is more suitable for simulating the motion of a large number of particles in fluid, particularly 
in the three-dimensional cases where particle collisions dominate. Dozens of times speedup can be 
expected in two-dimensional simulation 
and hundreds of times in three-dimensional simulation under the same particle and mesh number. 

Numerical simulations were carried out based on the LBM-PIBM-DEM scheme, our result of falling of single particle reveals 
that the settling velocity predicted by numerical simulation agrees well with the Stokes' law. Further multi-particle 
simulation results confirm that the LBM-PIBM-DEM scheme can capture the feature of the particulate flows in fluid and 
is a promising strategy for the solution of the particle-fluid interaction problems. By comparing two- and 
three-dimensional results, essential discrepancy was found due to the drawback of the two-dimensional assumption.
 Therefore, it can be concluded that the two-dimensional simulations may be good as a first and cheaper approach, 
 the three-dimensional simulations are necessary for an accurate description of the particle behaviors as well as 
 the flow patterns. From our three-dimensional results by PIBM, the sedimentation efficiency of particle is found to 
 increase with the decrease of initial porosity. 

Due to the fact that the calculation of the fluid-particle interaction force in the PIBM is simply based on the momentum 
conservation of the fluid particle, the LBM-PIBM-DEM scheme can be easily connected with other CFD solvers or Lagrangian 
particle tracking method where the conventional IBM works, e.g. with the direct numerical simulation~\cite{feng2014using}. 
However, during the simulations, 
we found that numerical instability may occur when the particle velocity is high, which seems to be a general weakness of 
the IBM family methods. For the sake of achieving validate results, the PIBM users are recommended to conduct a simplified 
case to compare with the analytical solutions/experimental observation to tune the LBM relaxation time, $\tau$, before using 
it in the multi-particle simulations. This practice is competent and has been widely used in LBM-DEM~\cite{Feng200520,Feng2009370}
and other simulations based on DEM~\cite{Tan2008975,Tan20091029}.  

\section*{Acknowledgments}

This work has been financially supported by the \textit{Ministerio de
Ciencia e Innovaci\'{o}n}, Spain (ENE2010-17801).
Hao Zhang would like to acknowledge the FI-AGAUR doctorate scolarship 
granted by the Secretaria d'Universitats i Recerca (SUR) del Departament d'Economia i Coneixement (ECO) 
de la Generalitat de Catalunya, and by the European Social Fund.
F. Xavier Trias would like to thank the financial support by the 
\textit{Ram\'{o}n y Cajal} postdoctoral contracts (RYC-2012- 11996)
by the \textit{Ministerio de Ciencia e Innovaci\'{o}n}. 

\section*{References}

\bibliography{./pibm}

\end{document}